\documentclass[conference]{IEEEtran}
\newcommand{\tabincell}[2]{\begin{tabular}{@{}#1@{}}#2\end{tabular}}

\usepackage{soul,listings,xcolor}

\usepackage{hyperref}

\usepackage{graphicx}

\lstnewenvironment{teX}[1][]
{\lstset{language=[LaTeX]TeX}\lstset{escapeinside={(*@}{@*)},
		numbers=left,numberstyle=\scriptsize,stepnumber=1,numbersep=5pt,
		breaklines=true,
		framesep=5pt,
		basicstyle=\scriptsize\ttfamily,
		showstringspaces=false,
		keywordstyle=\itshape\color{blue},
		stringstyle=\color{maroon},
		commentstyle=\color{black},
		rulecolor=\color{black},
		xleftmargin=0pt,
		xrightmargin=0pt,
		aboveskip=\medskipamount,
		belowskip=\medskipamount,
		backgroundcolor=\color{white}, #1
}}
{}

\usepackage{algorithm}
\usepackage{algorithmicx}
\usepackage{algpseudocode}
\usepackage{amsmath}

\usepackage{booktabs}

\usepackage{multirow}

\usepackage{threeparttable}

\usepackage[misc]{ifsym}

\begin{document}
\title{MobFuzz: Adaptive Multi-objective Optimization in Gray-box Fuzzing}

\author{\IEEEauthorblockN{Gen Zhang, Pengfei Wang\textsuperscript{\Letter}, Tai Yue, Xiangdong Kong, Shan Huang, Xu Zhou, Kai Lu\textsuperscript{\Letter}}
				\IEEEauthorblockA{College of Computer, National University of Defense Technology\\
				\{zhanggen, pfwang, yuetai17, kongxiangdong, huangshan12, zhouxu, kailu\}@nudt.edu.cn}}

\IEEEoverridecommandlockouts
\makeatletter\def\@IEEEpubidpullup{6.5\baselineskip}\makeatother
\IEEEpubid{\parbox{\columnwidth}{
		Network and Distributed Systems Security (NDSS) Symposium 2022\\
		24-28 April 2022, San Diego, CA, USA\\
		ISBN 1-891562-74-6\\
		https://dx.doi.org/10.14722/ndss.2022.24314\\
		www.ndss-symposium.org
	}
	\hspace{\columnsep}\makebox[\columnwidth]{}}

\maketitle

\begin{abstract}
Coverage-guided gray-box fuzzing (CGF) is an efficient software testing technique. There are usually multiple objectives to optimize in CGF. However, existing CGF methods cannot successfully find the optimal values for multiple objectives simultaneously. In this paper, we propose a gray-box fuzzer for multi-objective optimization (MOO) called MobFuzz. We model the multi-objective optimization process as a multi-player multi-armed bandit (MPMAB). First, it adaptively selects the objective combination that contains the most appropriate objectives for the current situation. Second, our model deals with the power schedule, which adaptively allocates energy to the seeds under the chosen objective combination. In MobFuzz, we propose an evolutionary algorithm called NIC to optimize our chosen objectives simultaneously without incurring additional performance overhead. To prove the effectiveness of MobFuzz, we conduct experiments on 12 real-world programs and the MAGMA data set. Experiment results show that multi-objective optimization in MobFuzz outperforms single-objective fuzzing in the baseline fuzzers. In contrast to them, MobFuzz can select the optimal objective combination and increase the values of multiple objectives up to 107\%, with at most a 55\% reduction in the energy consumption. Moreover, MobFuzz has up to 6\% more program coverage and finds 3x more unique bugs than the baseline fuzzers. The NIC algorithm has at least a {2x} improvement with a performance overhead of approximately 3\%.
\end{abstract}

\section{Introduction} \label{sec_intro}

Fuzz testing, or fuzzing, is one of the most successful search-based software testing approaches. Coverage-guided gray-box fuzzing (CGF), as an important variant of fuzzing, has recently received wide attention from researchers\cite{bohme2017coverage}. Essentially, CGF is an \textbf{optimization} problem \cite{sbst-first, mcminn2011search}. The key to an optimization approach is to search the input space to find optimal solutions and optimize the \textbf{objective}s. Optimizing the objectives means searching for the inputs that maximize or minimize the values of objectives \cite{sbst-achive}. 
In CGF, the most significant objective is code coverage, and the goal of CGF is to maximize coverage.

Single-objective optimization searches for the optimal solutions for only one objective. However, in real-world situations, more than one objective is required to be optimized simultaneously to solve difficult problems\cite{sbst-achive, sbst-recent}, such as detecting different kinds of bugs and improving fuzzing efficiency. Specifically, in different stages of the fuzzing process, these objectives should be adaptively selected and prioritized according to the testing scenario. For example, when testing code fragments of memory allocation, seeds regarding the objective of memory consumption should be prioritized; to break the embedded branch conditions, seeds with more satisfied comparison bytes should be an important objective. 
Thus, multi-objective optimization (MOO) is proposed to effectively study the balanced solutions with optimal trade-offs among multiple objectives \cite{deb2014multi,mcminn2011search, sbst-achive}.

Though coverage-guided fuzzers also consider objectives other than coverage in the searching process, existing gray-box fuzzers cannot really support multi-objective optimization. AFL \cite{afl}, for example, also searches for inputs with two other objectives, the execution time and input size. Favorable inputs (i.e., seeds) that have a smaller product of these two objectives ($\mathtt{speed*size}$)  are selected. Theoretically, in the search process, considering one solution at a time, e.g., the product of the objectives, may result in getting stuck in a local optimum and being unable to produce the global optimal solution \cite{mcminn2011search}. Some tools cannot coordinate multiple objectives simultaneously. When adding new objectives, old ones are discarded. For example, MemLock \cite{memlock} targets memory consumption bugs by choosing seeds with more memory consumption. It optimizes the objectives of both coverage and memory consumption. However, as a tool based on AFL, MemLock entirely removes the speed objective of AFL. This ignorance of multiple objectives clearly affects the execution speed of fuzzing according to our experiments.

Therefore, in reality, to properly optimize multiple objectives simultaneously in CGF, we have to overcome the following challenges: 1) \textbf{Conflict effects among different objectives.} During the long campaign of fuzzing, optimizing one objective may have a negative effect on another objective. For example, according to our experiments, pushing the number of satisfied comparison bytes to a large value to pass branch conditions will slow down the whole fuzzing process. This conflicting internal relationship among objectives requires us to properly coordinate different objectives in different stages to search for a global optimum solution. Thus, we conclude \textit{the adaptive selection of objective combination} as the first research point in this paper.

2) \textbf{Power schedule suitable for a multi-objective situation.} The power schedule of CGF is used to control the number of mutations and executions (i.e., amount of energy) on seeds \cite{afltech}, which directs the fuzzing process. Previous work on power schedule, such as AFLFast \cite{bohme2017coverage} and EcoFuzz \cite{ecofuzz}, aims to allocate an appropriate amount of energy based on the path discovery ability of seeds to save energy. However, under the multi-objective situation, the power schedule is required to combine with objective combination selection to control the energy allocation. Thus, we identify \textit{power schedule combined with objective combination selection} as the second research point in this paper.

3) \textbf{Reducing performance overhead introduced by multi-objective fuzzing.} Efficiency is an important metric in fuzzing. When taking multi-objectives into consideration, the objective combination selection, as well as the optimization on power schedule and mutation strategy all introduce additional overhead. For example, Cerebro\cite{cerebro} uses the idea of the Pareto frontier (i.e., the set of seeds with the optimal objective values) and non-dominated sorting \cite{nsga} to search for the optimum solutions in an evolutionary process. However, this process of Cerebro is executed only once in a fuzzing cycle. It keeps the fuzzer waiting for the final result and wastes precious CPU time. Additionally, a single run cannot produce the global optimal solution. Calculating the Pareto frontier and revealing the convergence usually require more than 100 iterations through the evolutionary process\cite{nsga}. Directly adopting this evolutionary procedure in fuzzing to find the optimal solution will bring significant performance overhead. Therefore, the third research point is to \textit{find the optimal results of the selected objectives without introducing additional performance overhead}.

To overcome the above challenges of gray-box fuzzing in MOO, in this paper, we propose MobFuzz. To deal with the adaptive selection of objective combinations, we model the process of CGF under multi-objectives as a multi-player multi-armed bandit (MPMAB) problem. The goal of the classic MAB model is to maximize the reward in finite trials by choosing the appropriate arms\cite{mab, ecofuzz}. We model the objective combinations as different players with their own goals to deal with the problem of combination selection. The best objective combination that has the maximum reward for the current fuzzing state is selected. To deal with the power schedule suitable for a multi-objective situation, we model the seeds as bandit's arms and classify the fuzzing states into exploration and exploitation. MobFuzz controls the number of mutations and executions on a seed through the adaptive power schedule. Using this model, MobFuzz allocates the appropriate amount of energy for seeds under the chosen combination to reach the optimal results and avoid a waste of energy. To address the third challenge, we propose an evolutionary algorithm called \underline{n}on-dominated sorting genetic algorithm \underline{i}n \underline{C}GF (\textbf{NIC}). It is designed based on the Pareto frontier and non-dominated sorting to search for optimal solutions in an evolutionary process. A new mutation strategy is designed by casting the objective combination to a corresponding mutation operator combination. The mutators that are more likely to increase the objective values are selected with a greater chance. In addition, we propose several methods, such as a shared seed pool, to reduce the performance overhead.

To prove the effectiveness of MobFuzz, we conduct a series of experiments on real-world target programs and the MAGMA data set. Experiment results show that multi-objective optimization in MobFuzz can outperform single-objective optimization in the baselines. Compared with the state-of-the-art fuzzers, such as MemLock and FuzzFactory, MobFuzz can optimize all the objectives simultaneously to reach the optimal values. Specifically, MobFuzz exceeds its competitors up to 107\% in the values of objectives. In addition, the results show the effectiveness of our MPMAB model and NIC algorithm. We reduce at most 55\% energy consumption, and NIC has at least a {2x} performance improvement compared with the baseline fuzzers with only 3.3\% performance overhead. Additionally, MobFuzz has at most 6\% more program coverage and finds 3x more bugs than the competitors. In conclusion, we make the following contributions in this paper:

\begin{itemize}
	\item We target the weakness of CGF in multi-objective optimization. We model the MOO in gray-box fuzzing as a multi-player MAB problem and adaptively select the objective combinations and allocate energy to seeds by the model.
	\item We propose the NIC algorithm in MobFuzz to solve the problems of existing fuzzers in finding the optimal results. NIC is integrated into the fuzzing loop and searches for the optimal objective values without introducing too much overhead.
	\item We implement MobFuzz and evaluate it with real-world programs and the MAGMA data set. The results demonstrate the effectiveness of multi-objective optimization in CGF.
\end{itemize}

\section{Background}

\subsection{CGF and Objectives}

As one of the most popular software testing techniques, fuzzing has recently developed rapidly, especially in the field of coverage-guided gray-box fuzzing\cite{2021Reinforcement, li2020unifuzz, tortoisefuzz, fairfuzz, ptfuzz}. Compared with the plain black-box fuzzing and complicated white-box fuzzing, the key of CGF is to maximize code coverage through lightweight instrumentation\cite{afltech}. As a representative of CGF, AFL exposed many security-critical vulnerabilities with these features\cite{aflbug}.

\textbf{Basic infrastructure. }CGF starts with selecting a seed from the seed pool according to its goal and then allocates energy to the selected seeds via the power schedule. The energy controls the number of mutations and executions to this seed. Next, the seed is mutated to generate test cases. When executing these test cases, CGF monitors whether they achieve new code coverage. Test cases that achieve new coverage will be saved as seeds to the seed pool. Later, CGF goes back to seed selection and starts the next round of fuzzing.

\textbf{Objectives in CGF. }In addition to code coverage, gray-box fuzzers usually need to maximize multiple objectives when maintaining the seed pool. For instance, AFL searches for seeds with a faster execution speed and smaller size. The product of them, i.e., $\mathtt{speed*size}$, is used to optimize the objectives. Seeds with a larger product result will be marked as \textbf{favored}, and they will be selected with a greater chance than the non-favored seeds.

However, existing coverage-guided fuzzers cannot support multi-objective optimization. Some fuzzers based on AFL disable the original speed objective in AFL when adding a new objective, such as MemLock \cite{memlock} and FuzzFactory \cite{fuzzfactory}; other solutions use simple algorithms to coordinate multiple objectives, which will get stuck at a local optimum \cite{afl}, or introduce unacceptable overhead \cite{cerebro}. Therefore, a solution that can optimize multiple objectives simultaneously without causing performance reduction is needed.

\subsection{MAB Problem}

\textbf{Exploration vs. Exploitation. } The trade-off between exploration and exploitation is an important concept in game theory\cite{mab}. Inspired by a player's choice to maximize the reward when playing a slot machine, the multi-armed bandit model is proposed to solve this problem\cite{bubeck2012regret}. According to the definition, a classic MAB model contains $\mathtt{N}$ parallel arms, and only one arm is selected each time. The expectation of reward for arm $\mathtt{i}$ ($\mathtt{i\in \{1, 2, ..., N\}}$) is defined as $\mathtt{R_i}$. The key to the MAB problem is to maximize the total reward within finite arm selections. However, before trying a certain arm, its reward is unknown. The process of \textbf{exploration} performs trials on an arm to acquire a more accurate calculation of its reward. When the rewards of all the arms are known, choosing the arm with the maximum reward is the process of \textbf{exploitation}. In conclusion, choosing the best arm (exploitation) will maximize the current total reward. In the long term, making trials on reward-unknown arms (exploration) will help reach a larger total reward \cite{patil2018greybox}. It is our goal to weigh the strengths and weaknesses between exploration and exploitation in the MAB problem.

\textbf{The inappropriateness of using MAB in CGF. }Previous work in fuzzing, such as EcoFuzz \cite{ecofuzz}, models seeds as arms in MAB and solves specific problems through this model, e.g., energy allocation. If we want to make improvements on MOO in CGF, the classic MAB model is inappropriate. First, as discussed above, there are two choices we need to make: objective combination selection and energy allocation. It is natural to consider the seeds as arms in the MAB model. However, a single-player MAB model is no longer applicable since different objective combinations stand for different players with their own goals. In other words, we need a multi-player MAB to model this process. Second, the reward of the classic MAB model is time-invariant, and the number of arms is constant\cite{ecofuzz}. However, as the fuzzing campaign continues, the rewards of the objective combinations and seeds will change accordingly. Additionally, the number of seeds is not constant during fuzzing. The above drawbacks of the classic MAB model drive us to propose a variant model of MAB that is suitable for multi-objectives in fuzzing.

\section{Adaptive Multi-objective Optimization}

\subsection{Overview}\label{sec_main}

	\begin{figure}[h]
		\centering
		\includegraphics[width=0.95\columnwidth]{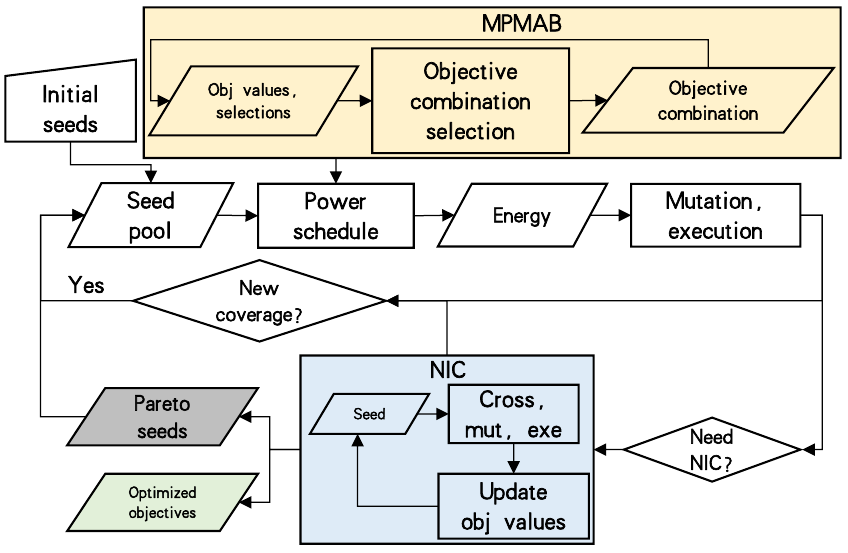}
		\caption{\footnotesize{The main fuzzing loop of MobFuzz. The sub-processes in different colors are our key approaches in MobFuzz.}}
		\label{fig_main}
	\end{figure}

As depicted in Figure \ref{fig_main}, MobFuzz is designed by adding two new modules to the classic fuzzing process: the \textbf{MPMAB} model and the \textbf{NIC} algorithm. The MPMAB model adaptively determines the objective combination and energy allocation. The NIC algorithm produces the optimal objective values through an evolutionary process without introducing additional performance overhead.

The basic procedure of MobFuzz contains the following steps. First, a seed is selected from the seed pool. Next, the MPMAB model determines the best objective combination under the current fuzzing status. Based on the chosen objective combination, MPMAB allocates different amounts of energy to seeds in exploration and exploitation states. Then, MobFuzz performs mutations and executions based on the allocated energy. At the same time, it monitors the objective values in the chosen combination. If a starting condition is met, the NIC algorithm is invoked. NIC is an evolutionary process. It updates the objective values to gradually approach the optimal solutions. Finally, the Pareto seeds with the optimal values are saved into the seed pool, and the next round of fuzzing begins.

\subsection{Multi-player Multi-armed Bandit Model}\label{sec_mlmab}

Our MPMAB model deals with two problems, including adaptively selecting objective combinations and allocating energy according to the chosen objective combination.

\subsubsection{MPMAB Model Overview}

\begin{table}[h]
	\setlength{\abovecaptionskip}{0cm}
	\caption{The names and definitions of variables in the MPMAB model}
	\label{table_var}
	\centering
	\resizebox{0.95\columnwidth}{!}{
		\begin{tabular}{ll}
			\bottomrule
			\bfseries Name & \bfseries Definition\\
			\toprule
			$t$ & \tabincell{c}{the ID of the current fuzzing round}\\
			$O_i$ & the $i$th objective\\
			$C_l$ & the $l$th combination\\
			$v^{t}_{\scriptscriptstyle O_i}$, $v^{t}_{\scriptscriptstyle C_l}$ & \tabincell{c}{the average value of objective $O_i$ or the objectives in $C_l$ in round $t$}\\
			$R$ & the reward\\
			$s_j$ & the $j$th seed\\
			$v_{\scriptscriptstyle O_i}(s_j)$ & \tabincell{c}{the value of objective $O_i$ after executing $s_j$}\\
			\bottomrule
		\end{tabular}
	}
\end{table}

\begin{figure}[h]
	\centering
	\includegraphics[width=0.8\columnwidth]{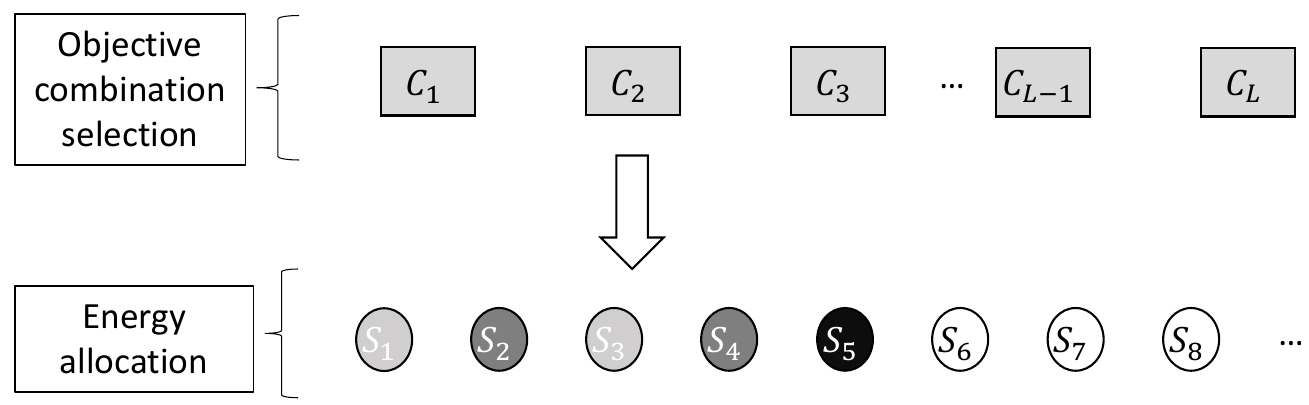}
	\caption{\footnotesize{Demonstration of the MPMAB model. The rectangles indicate objective combinations. The circles indicate seeds. The rewards of the colored shapes are known. The depth of color of the seeds indicates the amount of energy allocated.}}
	\label{fig_mlmab}
\end{figure}

Table \ref{table_var} shows the variables in our model. Figure \ref{fig_mlmab} shows the overview of our MPMAB model. First, our MPMAB model handles the problem of adaptive objective combination selection. Each combination stands for a player with his own goal when playing the slot machine. The number of objectives and the number of objective combinations are constant during fuzzing. For instance, if we have 3 objectives to optimize, there are 8 ($\mathtt{2^3}$) objective combinations in total. Since the number of combinations is constant, the reward of each combination can be captured through a pioneer stage. Then, we choose the best objective combination through our proposed algorithm (Section \ref{sec_com_sel}).

Second, MPMAB deals with the adaptive energy allocation of seeds under the chosen objective combination. During fuzzing, the number of seeds is increasing, and we cannot estimate the reward until we execute the seed. Therefore, the trade-off between focusing on reward-known seeds (exploitation) and trying reward-unknown seeds (exploration) is not as straightforward as in combination selection. To address this problem, we divide the fuzzing process into exploration and exploitation states. We can adaptively allocate energy in different states under different objective combinations (Section \ref{sec_pow_sch}).

\subsubsection{Objective Combination Selection}\label{sec_com_sel}

At the $\mathtt{t}$th minute, the ID of the current fuzzing round is $\mathtt{t}$. As the fuzzing campaign proceeds, we can obtain the average value of objective $\mathtt{O_i}$ during round $\mathtt{t}$, which is denoted as $\mathtt{v^{t}_{\scriptscriptstyle O_i}}$. In addition, we can calculate the average value of objective $\mathtt{O_i}$ in the previous $\mathtt{t}$ rounds (round 1, 2..., $\mathtt{t}$) as


	\begin{scriptsize}
	\begin{equation}
	\label{equ_ave_value}
		\overline{v}^{t}_{\scriptscriptstyle O_i}=\frac{\sum_{k=0}^{t}v^{k}_{\scriptscriptstyle O_i}}{t}
	\end{equation}
	\end{scriptsize}
Therefore, we define the reward of choosing objective $\mathtt{O_i}$ in round $\mathtt{t}$ as

	\begin{scriptsize}
	\begin{equation}
	\label{equ_reward}
		R(O_i, t) = t*(\frac{v^{t}_{\scriptscriptstyle O_i}}{\overline{v}^{t}_{\scriptscriptstyle O_i}}-\lambda*\frac{v^{t}_{\scriptscriptstyle O_0}}{\overline{v}^{t}_{\scriptscriptstyle O_0}})
	\end{equation}
	\end{scriptsize}
As we can see from the equation, $\mathtt{\frac{v^{t}_{\scriptscriptstyle O_i}}{\overline{v}^{t}_{\scriptscriptstyle O_i}}}$ denotes the ratio of the objective value in the current round and in the previous $\mathtt{t}$ rounds. When $\mathtt{v^{t}_{\scriptscriptstyle O_i}}$ is greater than $\mathtt{\overline{v}^{t}_{\scriptscriptstyle O_i}}$, the reward is large. This encourages selecting the objectives that increase rapidly. Additionally, we emphasize changes in late rounds \cite{aflhier} and multiple $\mathtt{t}$ in front of the reward. In addition, the speed of the fuzzing campaign is crucial to fuzz testing\cite{zeror, fullspeed, instrcr}; we add a penalty to objectives that slow down the process: $\mathtt{-\lambda*\frac{v^{t}_{\scriptscriptstyle O_0}}{\overline{v}^{t}_{\scriptscriptstyle O_0}}}$ (speed is the 0th objective).

When the number of objectives is $\mathtt{N}$, the number of objective combinations is $\mathtt{2^N}$. We deduce the reward for a combination $\mathtt{C_l}$ as

	\begin{scriptsize}
	\begin{equation}
		\label{equ_reward_com}
		R(C_l, t)=\frac{\sum_{O_i\in{C_l}}^{}{R(O_i, t)}}{L}+t*L
	\end{equation}
	\end{scriptsize}
$\mathtt{L}$ is the number of objectives in this combination. This reward consists of two parts. First, we use the average reward of the objectives. Second, $\mathtt{t*L}$ is added to reward combinations with more objectives.

Next, we can calculate a final score for the combinations to make decisions. UCB1 \cite{agrawal1995sample} is a classic answer to the MAB problem, and we calculate scores based on it as

	\begin{scriptsize}
	\begin{equation}
		\label{equ_usb}
		\begin{aligned}
		Score(C_l, t)&=\overline{R}(C_l, t)+U(C_l, t)\\
					 &=\frac{\sum_{k=0}^{t}{R(C_l, k)}}{t}+\gamma*\sqrt{\frac{ln(\sum_{C_l\in{C}}^{}{n_l})}{n_l}}
		\end{aligned}
	\end{equation}
	\end{scriptsize}
$\mathtt{C}$ denotes all the objective combinations. The score consists of two parts. $\mathtt{\overline{R}(C_l, t)}$ is the average reward of $\mathtt{C_l}$ in previous $\mathtt{t}$ rounds, which gives combinations with greater historical rewards higher scores (exploitation). $\mathtt{U(C_l, t)}$ is the upper confidence bound of $\mathtt{C_l}$, and it adds greater scores to combinations with smaller $\mathtt{n_l}$ values (the number of times the combination is selected), which is exploration. At the beginning of fuzzing, we go through a \textbf{pioneer stage}, in which each objective combination is selected once. After this stage, the $\mathtt{n_l}$ value of each combination will be 1. Next, at the end of each round of fuzzing, we calculate the score of each combination and choose the one with the maximum score as the objective combination for the next round. Moreover, $\mathtt{\gamma}$ is a key parameter in UCB1 that controls the balance between exploration and exploitation, and we will discuss this parameter in Section \ref{sec_eva}.

\subsubsection{Power Schedule}\label{sec_pow_sch}

Once we determine the objective combination in this round of fuzzing, our MPMAB model goes to adaptive energy allocation under the chosen objective combination. The number of seeds (arms) increases as the fuzzing campaign continues, and we cannot reuse a pioneer stage to get the reward to start the model. As mentioned above, our key challenge is to balance exploration (trying reward-unknown seeds) and exploitation (choosing the seed with the maximum reward). Our goal is to adaptively assign energy under the chosen objective combination in different fuzzing states. However, according to the related research in energy allocation of CGF including \cite{bohme2017coverage, ecofuzz, bohme2020boosting}, there is no previous work about allocating energy in the situation of multi-objective optimization. Based on this background, we first define the average energy to reach a certain objective value as

	\begin{scriptsize}
	\begin{equation}
		\label{equ_ave_eng}
		\overline{E}^t_{\scriptscriptstyle O_i}=\frac{Execs(t)}{\overline{v}^{t}_{\scriptscriptstyle O_i}}
	\end{equation}
	\end{scriptsize}
This is the quotient of the number of executions and the value of an objective in $\mathtt{t}$ rounds. We consider it the \textbf{minimum} energy required to increase the objective value.

Likewise, we can deduce the average energy of $\mathtt{C_l}$ in $\mathtt{t}$ rounds as

	\begin{scriptsize}
	\begin{equation}
		\label{equ_ave_eng_cl}
		\overline{E}^t_{\scriptscriptstyle C_l}=\frac{\sum_{O_i\in{C_l}}^{}{\overline{E}^t_{\scriptscriptstyle O_i}}}{L}
	\end{equation}
	\end{scriptsize}
Next, we divide the fuzzing states into different states: \textbf{Exploration state. }This state implies that there are currently reward-unknown seeds, and we need to try as many new seeds as possible. We allocate the minimum energy to seeds in this state as

	\begin{scriptsize}
	\begin{equation}
		\label{equ_ra_eng}
		E(s_j)=\overline{E}^t_{\scriptscriptstyle C_l}
	\end{equation}
	\end{scriptsize}
This amount of energy has two features: 1) It remains very small. 2) Although it is small, it is the minimum expected energy needed to reach greater objective values. Therefore, we assign this amount of energy to seeds in this state.

\textbf{Exploitation state. }In this state, all the rewards of seeds are known, and it is rational to choose the seed with the maximum reward. We define the energy in this state as 

	\begin{scriptsize}
	\begin{equation}
	\label{equ_loita_eng}
	E(s_j)=\overline{E}^t_{\scriptscriptstyle \scriptscriptstyle C_l}*(\frac{v_{\scriptscriptstyle C_l}(s_j)}{\overline{v}^{t}_{\scriptscriptstyle C_l}}+\sum_{O_i\in{C_l}}^{}{is\_max(v_{\scriptscriptstyle O_i}(s_j))})
	\end{equation}
	\end{scriptsize}
We can see that Equation \ref{equ_loita_eng} is based on the energy in the exploration state in Equation \ref{equ_ra_eng}. In this equation, $\mathtt{\frac{v_{\scriptscriptstyle C_l}(s_j)}{\overline{v}^{t}_{\scriptscriptstyle C_l}}}$ is the ratio between the objective value of executing seed $\mathtt{s_j}$ and the average value. When the objective value of executing the seed ($\mathtt{v_{\scriptscriptstyle C_l}(s_j)}$) is greater than the average value ($\mathtt{\overline{v}^{t}_{\scriptscriptstyle C_l}}$), we allocate more energy to encourage it and vice versa. Additionally, $\mathtt{is\_max()}$ returns 1 if executing this seed reaches the maximum value of a certain objective and returns 0 otherwise. If executing this seed reaches the maximum value of a certain objective, we add a bonus based on this function ($\mathtt{is\_max(v_{\scriptscriptstyle O_i}(s_j))}$) to the allocated energy.

\subsection{NIC Algorithm}\label{sec_nic}

\subsubsection{NIC Overview} \label{sec_nic_overview}
\begin{algorithm}
	\caption{The MobFuzz algorithm.}
	\label{algo_main}
	\scriptsize
	\begin{algorithmic}[1]
		\Require Initial seeds $S$, Objectives $O$
		\State $Q=S$ \label{alg_main_initial_seed}
		\While{$cur\_time < TIME\_OUT$}
		\If{$cur\_time-prev\_time >= 1{\,\,}min$} \label{alg_main_obj_sel_start}
		\State \textsl{/* combination selection: {obj\_com\_sel(obj\_val, obj\_sel, t)} */}
		\State $Score=\overline{R}(C_l, t)+U(C_l, t)$
		\State $C_l=arg\_max(Score)$
		\State \textsl{/* \textsl{end of combination selection} */} 
		\State $prev\_time=cur\_time$
		\EndIf \label{alg_main_obj_sel_end}
		\State \textsl{/* power schedule: {pow\_sch(Cl, obj\_val, t)} */} \label{alg_main_pow_sch_start}
		\If{$state==Exploration$} \textsl{/* current state of fuzzing */}
		\State $energy = \overline{E}^t_{\scriptscriptstyle C_l}$
		\EndIf
		\If{$state==Exploitation$}
		\State $energy = \overline{E}^t_{\scriptscriptstyle C_l}*(\frac{v_{\scriptscriptstyle C_l}(s_j)}{\overline{v}^{t}_{\scriptscriptstyle C_l}}+\sum_{O_i\in{C_l}}^{}{is\_max(v_{\scriptscriptstyle O_i}(s_j))})$
		\EndIf
		\State \textsl{/* end of power schedule */}\label{alg_main_pow_sch_end}
		\For{$i = 0 \to energy$} \textsl{/* energy assigned for each seed */}
		\State mut\_exe($s$) \label{alg_main_mut} \textsl{/* seed s is selected by AFL mechanism */}
		\If{$new\_cov(s)==True$} \label{alg_main_cov_start}
		\State $Q = Q+s$
		\EndIf \label{alg_main_cov_end}
		\If{$need\_nic==true$} \label{alg_main_need_nic}
		\State \textsl{/* NIC: {NIC(s', Cl, T)} */} \label{alg_main_nic_start}
		\For{$j = 0 \to T$} \label{alg_main_25}
		\State cross\_mut\_exe($s'$) \label{alg_main_26} \textsl{/* s' is the seed in NIC */}
		\State update($O_M$) \label{alg_main_27}
		\If{$new\_cov(s')==True$} \label{alg_main_28}
		\State $Q = Q+s'$ \label{alg_main_nic_q}
		\EndIf \label{alg_main_30}
		\EndFor
		\State return $Pareto$ \label{alg_main_32} \textsl{/* Pareto frontier */}
		\State \textsl{/* end of NIC */}\label{alg_main_nic_end}
		\EndIf
		\State $Q = Q+Pareto$ \label{alg_main_35}
		\EndFor
		\EndWhile
		\Ensure Pareto seeds with optimized objectives $O_M$
	\end{algorithmic}
\end{algorithm}

To optimize the objectives in the chosen combination and find the optimal result without additional performance overhead, we propose our NIC algorithm. It is an evolutionary algorithm designed for multi-objective optimization in MobFuzz. The basic process of NIC is as follows: First, an initial population of seeds with a scale of $\mathtt{N}$ is selected. Next, the offspring seeds are obtained with crossover and mutation among the initial population. Second, we execute the target program with each seed in the population and obtain related information (Line \ref{alg_main_26} in Algorithm \ref{algo_main}), e.g., coverage information and objective values. From the second generation onwards, the parent population and the offspring are combined to perform non-dominated sorting \cite{nsga}. According to the non-dominated relationship of seeds, the seeds with updated objective values (Line \ref{alg_main_27}) are selected to form a new parent population with a scale of $\mathtt{N}$. Additionally, the coverage information helps update the seed pool (Lines \ref{alg_main_28} - \ref{alg_main_30}). Finally, the new offspring seeds are generated by the crossover and mutation among the new parent population. This process repeats until the pre-defined number of iterations is met, and NIC outputs the Pareto frontier, which contains the seeds with the optimal objective values (Line \ref{alg_main_32}).

\subsubsection{Detailed Techniques of NIC}
\textbf{Adaptive population size. }The scale of the population is a crucial factor in the NIC algorithm. A small population lacks diversity and cannot produce the optimal result. A large population requires more computing resources in the evolutionary process. Therefore, we need to strike a balance in the size of the population before starting NIC. Through extensive testing, we determine that 10\% of the number of seeds should be an appropriate value for the population size. Before we start NIC each time, we randomly select 10\% of the seeds from the seed pool to form the initial population, and they go into the evolutionary process.

\textbf{Co-mutation operators with AFL. }We propose 3 techniques for efficient mutation in NIC. First, we integrate the fuzzing-effective operators into NIC, e.g., replacing four bytes with a boundary value of the integer type. Moreover, traditional evolutionary or genetic algorithms retain both offspring to maintain the \textbf{diversity} of the population. To address this problem, in NIC, two parent seeds go through crossover and mutation, and we keep both of the generated offspring. Finally, since AFL selects different mutation operators and positions to mutate with equal probability, it cannot highlight the importance of different operators and positions. In addition, in the situation of multi-objective optimization, we need to connect objectives to a specific or some mutation operators. In other words, an objective combination should have a corresponding mutation operator combination, in which the operators should be given a higher probability. To handle these issues, in NIC, we record the number of times the operators and positions can increase the objective values in the chosen objective combination. The ones that are more likely to increase the objective values are selected with a greater probability. In this way, we can get the best mutation operators or positions for a chosen combination.

\textbf{Reducing performance overhead.} The evolutionary or genetic algorithms usually iterate for more than 100 generations. If we directly inserted this process into the fuzzing loop, the performance overhead would be unacceptable. We propose several methods to handle this issue. For example, as described in Section \ref{sec_nic_overview}, when we crossover and mutate the parent population to generate the offspring, we add the process of the new coverage monitor. Seeds generated during NIC that cover new code can also be saved to the seed pool of the main fuzzing loop. Techniques such as this shared seed pool indicate that NIC is no longer independent of the main loop. They are integrated instead. In this way, the overhead to produce the optimal results through the iterations is removed in NIC. In addition, we add a starting condition for NIC. In our design, when we monitor a decrease in the values of the objectives, we start the NIC algorithm to optimize and increase them. Specifically, we record the objective values every minute. When the gradient of two continuous values is less than a threshold, the NIC algorithm is started. According to our preliminary experiments, when the threshold is -0.15, it is the configuration to start NIC to achieve the best performance. Additionally, NIC will execute for a pre-defined number of iterations (Line \ref{alg_main_25} in Algorithm \ref{algo_main}).


\subsection{Working Flow}

In Algorithm \ref{algo_main}, the inputs of MobFuzz are the initial seeds $\mathtt{S}$ and the objectives $\mathtt{O}$ we want to optimize. The seed pool $\mathtt{Q}$ is initialized to the user-provided seeds on Line \ref{alg_main_initial_seed}. Within the configured timeout period of fuzzing ($\mathtt{TIME\_OUT}$), MobFuzz continues to fuzz the target program.

To begin, we need to determine the time interval to select the objective combination for the next round. Based on our preliminary evaluation, we take 1 minute as the time interval to make selections. The 24-hour fuzzing campaign is thus divided into 1440 ($\mathtt{\frac{24*60}{1}}$) rounds. Each round $\mathtt{t}$ of fuzzing lasts for 1 minute. At the end of round $\mathtt{t}$, we need to select the objective combination for the next round. As shown on Lines \ref{alg_main_obj_sel_start} - \ref{alg_main_obj_sel_end}, if the time interval exceeds 1 minute, we will adaptively select objective combinations with MPMAB in function $\mathtt{obj\_com\_sel()}$. The arguments of this function include $\mathtt{obj\_val}$ (the values of the objectives), $\mathtt{obj\_sel}$ (the numbers of selections of the objectives), and $\mathtt{t}$ (the current fuzzing round). This function outputs the selected objective combination $\mathtt{C_l}$. When a combination is determined, in this time interval, we will optimize the objectives in this chosen combination. Section \ref{sec_com_sel} shows this procedure in detail.

On Lines \ref{alg_main_pow_sch_start} - \ref{alg_main_pow_sch_end} in function $\mathtt{pow\_sch()}$, we monitor the current state (exploration or exploitation) of fuzzing and assign energy to seeds based on the chosen objective combination. The arguments are listed on Line \ref{alg_main_pow_sch_start}. The output is the energy according to Equation \ref{equ_ra_eng} and \ref{equ_loita_eng} in Section \ref{sec_pow_sch}. The amount of energy determines the number of mutations and executions on a seed (Line \ref{alg_main_mut}). Specifically, we focus on the power schedule. Therefore, we inherit the seed selection mechanism of AFL. After the mutations and executions ($\mathtt{mut\_exe(s)}$), we save seeds that bring new code coverage (Lines \ref{alg_main_cov_start} - \ref{alg_main_cov_end}). Lines \ref{alg_main_nic_start} - \ref{alg_main_nic_end} show the workflow of NIC. The arguments of $\mathtt{NIC()}$ include $\mathtt{s'}$ (the selected seeds from the pool), $\mathtt{C_l}$ (the chosen objective combination), and $\mathtt{T}$ (the number of iterations of NIC). Line \ref{alg_main_26} shows the mutations and executions of NIC. The values of objectives $\mathtt{O_M}$ are updated on Line \ref{alg_main_27}. Additionally, the NIC algorithm goes through the evolutionary process with a shared seed pool (Line \ref{alg_main_nic_q}) and optimizes the objective values. The output of NIC is the Pareto frontier \cite{nsga} which is the set of seeds with the optimal objective values. We add the Pareto frontier as seeds to the seed pool on Line \ref{alg_main_35}. In conclusion, NIC has the following functionalities: 1) It outputs the optimal objective values of the selected objectives. 2) NIC outputs the Pareto frontier, and these seeds are saved to the shared seed pool. 3) At the same time, it saves seeds that bring new coverage.


\section{Implementation}

We implement MobFuzz based on AFL. We modify the instrumentation code in LLVM to record the amount of stack memory consumption and number satisfied comparison bytes, respectively. The MPMAB model and the NIC algorithm are implemented separately in $\mathtt{afl}$-$\mathtt{fuzz.c}$, and they contain {1.5k} lines of code in total. Additionally, we modify the main fuzzing loop to interact with MPMAB and NIC. We replace the original power schedule with the MPMAB schedule, and we add code to check if the starting condition of NIC is met. Specifically, when the gradient of two continuous objective values is less than a threshold, NIC will be started. These modifications contain about {0.5k} lines of code.


\section{Evaluation} \label{sec_eva}

In our evaluation, we answer the following research questions:
\begin{itemize}
	\item \textbf{RQ1. }How does multi-objective optimization in MobFuzz perform compared with single-objective optimization in the baseline fuzzers?
	\item \textbf{RQ2. }How does the objective combination selection adapt in the fuzzing process?
	\item \textbf{RQ3. }How does our power schedule perform compared with the baseline fuzzers under the chosen objective combination?
	\item \textbf{RQ4. }Does NIC optimize the objective values without introducing additional performance overhead?
\end{itemize}

\subsection{Setup}

\textbf{Target programs to test. }We test 12 real-world programs in total. They include programs of various purposes, e.g., image processing ($\mathtt{tiff2pdf}$). Table \ref{table_target} shows the basic information of these target programs. They are collected from state-of-the-art papers and these papers are listed in Table \ref{table_target_use} in Appendix. We believe collecting programs in this way can ensure persuasiveness and representativeness.


\begin{table}[h]
	\setlength{\abovecaptionskip}{0cm}
	\caption{Target programs}
	\label{table_target}
	\centering
	\resizebox{0.7\columnwidth}{!}{
		\begin{tabular}{ccc}
			\bottomrule
			\bfseries Targets & \bfseries Version & \textbf{Format}\\
			\toprule
			avconv -y -i @@ -f null & libav-12.3 & mp4\\
			exiv2 @@ /dev/null & exiv2-0.27 & jpeg\\
			infototap @@ & ncurses-6.1 & txt\\
			mp42aac @@ a.aac & Bento4-1.5.1-628 & mp4\\
			\tabincell{c}{mp4tag --show-tags --list-\\symbols --list-keys @@} & Bento4-1.5.1-628 & mp4\\
			nm -C @@ & Binutils-2.30 & elf\\
			\tabincell{c}{podofopdfinfo @@} & podofo-0.9.6 & pdf\\
			readelf -a @@ & Binutils-2.30 & elf\\
			tiff2pdf @@ & libtiff-4.0.7 & tiff\\
			tiff2ps @@ & libtiff-4.0.7 & tiff\\
			\tabincell{c}{podofotextextraxt @@} & podofo-0.9.6 & pdf\\
			xmllint @@ & libxml-2.98 & xml\\
			\bottomrule
		\end{tabular}
	}
\end{table}

	\begin{table*}
		\setlength{\abovecaptionskip}{0cm}
		\caption{Evaluation of MobFuzz regarding different objectives}
		\label{table_objective}
		\centering
		\resizebox{0.9\textwidth}{!}{
			\begin{threeparttable}
			\begin{tabular}{cccccccccc}
				\bottomrule
				\multirow{2}*{\textbf{Targets}} & \multicolumn{3}{c}{\textbf{Execution speed}} & \multicolumn{3}{c}{\textbf{Stack memory consumption}} & \multicolumn{3}{c}{\textbf{Satisfied comparison bytes}} \\
				\cline{2-10}
				\bfseries ~ & \bfseries MobFuzz & \textbf{AFL} & \textbf{p value/$\mathtt{\hat{A}_{12}}$} & \textbf{MobFuzz} & \textbf{MemLock} & \textbf{p value/$\mathtt{\hat{A}_{12}}$} & \textbf{MobFuzz} & \textbf{FuzzFactory} & \textbf{p value/$\mathtt{\hat{A}_{12}}$} \\
				\toprule
				avconv & \textbf{85.60}\tnote{1} & 83.09 & 0.02/0.77 & \textbf{58704.00} & 37846.40 & 0.01/0.75 & $\textbf{2.88}\mathbf{*10^8}$ & 1.51$*10^8$ & $<10^{-4}$/1.00 \\
				exiv2 & \textbf{671.41} & 372.58 & $<10^{-4}$/1.00 & \textbf{1.55}$\mathbf{*10^6}$ & 1.21$*10^6$ & 0.01/0.80 & \textbf{4.66}$\mathbf{*10^7}$ & 2.24$*10^7$ & $<10^{-3}$/0.92\\
				infotocap & \textbf{386.86} & 292.55 & $<10^{-2}$/0.86 & \textbf{22122.40} & 15852.00 & 0.12/0.60 & \textbf{2.05}$\mathbf{*10^8}$ & 1.70$*10^8$ & $<10^{-2}$/0.90 \\
				mp42aac & \textbf{1688.83} & 1666.99 & 0.40/0.54 & \textbf{2.03}$\mathbf{*10^5}$ & 1.81$*10^5$ & 0.02/0.77 & \textbf{8.34}$\mathbf{*10^7}$ & 4.20$*10^7$ & $<10^{-3}$/0.93 \\
				mp4tag & \textbf{1452.70} & 1381.56 & 0.08/0.69 & \textbf{2.25}$\mathbf{*10^5}$ & 1.98$*10^5$ & 0.01/0.79 & \textbf{7.67}$\mathbf{*10^7}$ & 4.23$*10^7$ & 0.01/0.80\\
				nm & \textbf{1426.29} & 972.73 & $<10^{-3}$/0.94 & \textbf{8.38}$\mathbf{*10^6}$ & 7.17$*10^6$ & 0.09/0.68 & \textbf{1.88}$\mathbf{*10^8}$ & 6.57$*10^7$ & $<10^{-4}$/1.00\\
				pdfinfo & \textbf{875.96} & 642.15 & $<10^{-4}$/1.00 & 3660.00 & 3660.00 & -/0.50 & \textbf{1.14}$\mathbf{*10^7}$ & 7.29$*10^6$ & $<10^{-4}$/1.00\\
				readelf & \textbf{1361.56} & 1056.57 & 0.05/0.72 & \textbf{2663.60} & 2382.40 & 0.43/0.52 & \textbf{3.08}$\mathbf{*10^8}$ & 7.83$*10^7$ & $<10^{-4}$/1.00\\
				tiff2pdf & \textbf{1895.09} & 1670.86 & $<10^{-4}$/1.00 & 1588.00 & 1588.00 & -/0.50 & \textbf{4.38}$\mathbf{*10^7}$ & 3.86$*10^7$ & 0.25/0.49\\
				tiff2ps & \textbf{2050.05} & 1750.97 & $<10^{-2}$/0.87 & 1308.00 & 1308.00 & -/0.50 & \textbf{9.16}$\mathbf{*10^6}$ & 8.48$*10^6$ & 0.03/0.74\\
				txtext & \textbf{855.39} & 505.29 & $<10^{-4}$/1.00 & \textbf{3876.10} & 3676.00 & 0.03/0.75 & \textbf{1.17}$\mathbf{*10^7}$ & 1.12$*10^7$ & 0.05/0.67\\
				xmllint & \textbf{1020.47} & 746.53 & $<10^{-4}$/1.00 & \textbf{64112.80} & 64085.60 & 0.03/0.72 & \textbf{1.65}$\mathbf{*10^8}$ & 6.65$*10^7$ & $<10^{-4}$/1.00\\
				\toprule
				Average & \textbf{1147.51} & 928.48({\footnotesize{+23.6\%}})\tnote{2} & {0.04/0.87} & \textbf{8.72}$\mathbf{*10^5}$ & 7.40$*10^5$({\footnotesize{+17.7\%}}) & {0.04/0.66} & \textbf{9.60}$\mathbf{*10^7}$ & {4.62$*10^7$}({\footnotesize{+107.9\%}}) & {0.02/0.87}\\
				\bottomrule
			\end{tabular}
			\begin{tablenotes}
				\footnotesize
				\item[1] Greater values are better. $^2$ The percentages in the brackets of the last line denote the increase in contrast to the baseline fuzzers.
			\end{tablenotes}
		\end{threeparttable}
		}
	\end{table*}

\textbf{Baseline fuzzers to compare. }AFL\cite{afl}, MemLock\cite{memlock}, and FuzzFactory\cite{fuzzfactory} are used in our evaluation to test real-world programs. According to our discussion in Introduction, they are chosen because we can compare the multi-objective optimization of MobFuzz with the single-objective optimization of them. In addition, we choose 3 objectives as our evaluation metrics, including speed of execution (AFL), stack memory consumption (MemLock), and number of satisfied comparison bytes (FuzzFactory)\footnote{SP or Speed  denotes execution speed, ST or Stack denotes stack memory consumption, and CM or Cmp denotes number of satisfied comparison bytes.}. These fuzzers use consistent settings: deterministic and havoc.

We use seeds in the $\mathtt{testcase}$ directory provided by AFL as the initial seeds. Our evaluations are conducted on a server for 10 times and 24 hours.


\subsection{Effectiveness of Multi-objective Optimization}

\subsubsection{Results of Objective Values}Table \ref{table_objective} shows the average objective values of 10 repeated runs. The p values and $\mathtt{\hat{A}_{12}}$ values are also listed in the table. The values of MobFuzz are greater than (33 out of 36) or equal to (3 out of 36) the compared values in all the comparisons. Among them, 32 pairs of comparisons show a statistically significant difference ($p < 0.05$ or $\mathtt{\hat{A}_{12}} > 0.5$). Specifically, we have 10 comparisons that reach a p value less than $\mathtt{10^{-4}}$ and an $\mathtt{\hat{A}_{12}}$ value of 1.0. For example, in xmllint the satisfied comparison bytes of MobFuzz and FuzzFactory are $\mathtt{1.65*10^8}$ and $\mathtt{6.65*10^7}$, respectively. MobFuzz is approximately 2x better than that of FuzzFactory. In the $\mathtt{Average}$ row, we can see that the average value of MobFuzz is greater than that of the baseline fuzzers. In the value of satisfied comparison bytes, we even achieve more than a 100\% increase in contrast to FuzzFactory.

MOO is designed to accomplish the task of generating the optimal values for all the objectives. The reason for the improvement against MemLock and FuzzFactory is the NIC algorithm that helps MobFuzz reach the optimal values. NIC keeps looking for the Pareto seeds in an evolutionary process. Every iteration is closer to the optimal values. The final result is closest to the optimal values. MemLock and FuzzFactory have no such mechanism to reach the optimal values for the objectives.

\begin{table*}
	\setlength{\abovecaptionskip}{0cm}
	\caption{Evaluation based on branch coverage and number of unique bugs}
	\label{table_path}
	\centering
	\resizebox{0.7\textwidth}{!}{
		\begin{threeparttable}
		\begin{tabular}{ccccccccc}
			\bottomrule
			\multirow{2}*{\textbf{Targets}} & \multicolumn{4}{c}{\textbf{Number of edges}} & \multicolumn{4}{c}{\textbf{Number of unique bugs}}\\
			\cline{2-9}
			\bfseries ~ & \bfseries MobFuzz & \textbf{AFL} & \textbf{MemLock} & \textbf{FuzzFactory} & \bfseries MobFuzz & \textbf{AFL} & \textbf{MemLock} & \textbf{FuzzFactory} \\
			\toprule
			avconv & \textbf{24366.3}\tnote{1} & 22692.1 & 23750.8 & 21676.3 & 0.0 & 0.0 & 0.0 & 0.0 \\
			exiv2 & 11890.6 & 11144.7 & \textbf{11900.1} & 11769.1 & 0.0 & 0.0 & 0.0 & 0.0 \\
			infotocap & 2319.4 & \textbf{2421.3} & 2351.0 & 2213.9 & \textbf{2.8} & 0.0 & 0.0 & 0.0 \\
			mp42aac & \textbf{2890.6} & 2597.6 & 2722.7 & 2747.7 & \textbf{2.4} & 2.0 & 1.0 & 0.0 \\
			mp4tag & \textbf{3081.9} & 2733.4 & 2949.7 & 2875.8 & \textbf{2.5} & 0.0 & 1.0 & 0.0 \\
			nm & \textbf{7182.1} & 7077.5 & 7041.5 & 6923.6 & 0.0 & 0.0 & 0.0 & 0.0 \\
			pdfinfo & \textbf{2988.3} & 2608.3 & 2608.3 & 2608.3 & 0.0 & 0.0 & 0.0 & 0.0 \\
			readelf & 11444.4 & 10720.5 & \textbf{11180.4} & 11571.9 & 0.0 & 0.0 & 0.0 & 0.0 \\
			tiff2pdf & 1471.6 & \textbf{1474.6} & 1469.2 & 1471.0 & \textbf{1.0} & 0.0 & 0.0 & 0.0 \\
			tiff2ps & \textbf{451.0} & 425.4 & 425.4 & 426.0 & \textbf{1.0} & 0.0 & 0.0 & 0.0 \\
			txtext & \textbf{2789.1} & 2687.0 & 2532.0 & 2599.0 & 0.0 & 0.0 & 0.0 & 0.0 \\
			xmllint & \textbf{6072.8} & 5878.0 & 5944.7 & 5645.0 & 0.0 & 0.0 & 0.0 & 0.0 \\
			\toprule
			Average & \textbf{6412.3} & {6038.4({6\%})}\tnote{2} & {6239.7({3\%})} & {6044.0({6\%})} & {0.8} & {0.2({300\%})} & 0.2({300\%}) & 0({$\infty$\%}) \\
			\bottomrule
		\end{tabular}
		\begin{tablenotes}
			\footnotesize
			\item[1] Greater values are better. $^2$ The percentages in the brackets of the last line denote the increase in contrast to the baseline fuzzers.
		\end{tablenotes}
	\end{threeparttable}
	}
\end{table*}

\subsubsection{Branch Coverage and Unique Bugs}

Table \ref{table_path} shows the branch coverage and number of unique bugs found by the fuzzers. Branch coverage is the recommended coverage metric in evaluating fuzzing \cite{klees2018evaluating}. In regard to the number of branches, MobFuzz outperforms the baseline fuzzers in 30 out of the 36 comparisons. On average, it can find 6\% more branches at most than AFL and FuzzFactory. The unique bugs are manually collected from the unique crashes with the help of AddressSanitizer (ASAN) and gnu debugger (GDB). In the columns of unique bugs, despite the all-zero target programs, MobFuzz outperforms other fuzzers in all the 15 comparisons. The average value of MobFuzz is greater than the competitors, with 3x more bugs compared with AFL and MemLock.

We can see that MobFuzz outperforms the 3 competitors in the comparisons. The reason for this is that MobFuzz has better multi-objective optimization ability than other fuzzers. First, we retain a relatively fast speed in MobFuzz during the fuzzing campaign. This gives MobFuzz more chances to have more coverage and find more unique bugs. Additionally, MobFuzz has greater values in stack memory consumption than the competitors, which can lead to more bugs of the target programs. Third, the number of satisfied comparison bytes of MobFuzz is greater than other fuzzers. Satisfying more comparisons helps in exploring more program branches. Based on these reasons, MobFuzz can have more program coverage and find more unique bugs than the baseline fuzzers.

$\bullet$ \textit{Answer to RQ1: Multi-objective optimization in MobFuzz outperforms every single-objective optimization simultaneously in the baseline fuzzers.}

\subsection{Objective Combination Selection}

	\begin{figure}[htb]
		\centering
		\includegraphics[width=\columnwidth]{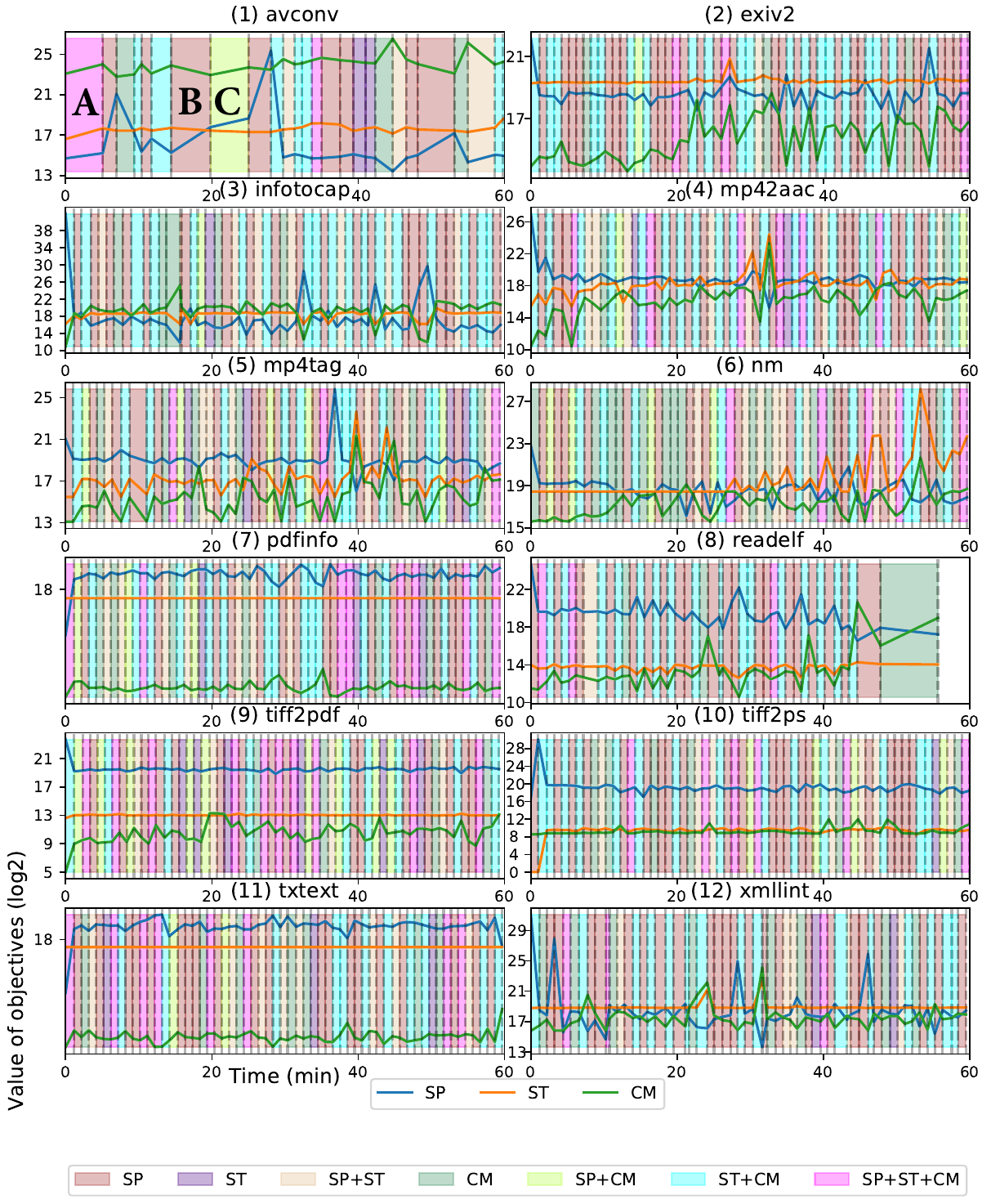}
		\caption{\footnotesize{Values of the objectives and the selected objective combinations within 60 minutes. The 3 lines indicate the values of the objectives. The background colors represent the selected objective combinations within each time interval.}}
		\label{fig_ave_data}
	\end{figure}
	
1) We set 1 minute as the time interval to make selections. During the 24 hours of the fuzzing campaign, 1,440 selections are made by the MPMAB model in total (possibly less because the target program may still be executing at the end of the time interval). Figure \ref{fig_ave_data} shows how the selected combination affects the values of objectives within each time interval. The background colors show the selected objective combination during this minute. The lines show the values of each objective. We take the result of $\mathtt{avconv}$ as an example. We mark points $\mathtt{A}$, $\mathtt{B}$ and $\mathtt{C}$ in the figure of $\mathtt{avconv}$. First, point $\mathtt{A}$ is marked to prove the ability of MobFuzz to optimize multiple objectives simultaneously. In this time interval, the chosen combination is $\mathtt{speed/stack/cmp}$. As we can see from the figure, in this round, all the values of objectives increase. Next, at point $\mathtt{B}$, $\mathtt{stack/cmp}$ is selected. We note that as stack and cmp increase, speed decreases, which demonstrates opposite effects between objectives. Finally, point $\mathtt{C}$ shows our correction to the slowdown of fuzzing. In this time interval, a penalty is added to the stack and cmp objectives because of point $\mathtt{B}$, and $\mathtt{speed}$ is selected to increase the execution speed of fuzzing.

	\begin{table}[htb]
		\setlength{\abovecaptionskip}{0cm}
		\caption{Percentages of the chosen objective combinations}
		\label{table_com_sel}
		\centering
		\resizebox{0.8\columnwidth}{!}{
			\begin{tabular}{cccccccc}
				\bottomrule
				\bfseries Targets & \bfseries SP & \textbf{ST/SP/CM} & \textbf{ST/CM} & \textbf{SP/CM} & \textbf{CM} & \textbf{SP/ST} & \textbf{ST}\\
				\toprule
				avconv & \textbf{63.4\%} & 3.8\% & 5.7\% & 6.5\% & 6.3\% & 8.5\% & 5.8\%\\
				exiv2 & \textbf{86.1\%} & 4.6\% & 2.1\% & 2.1\% & 1.7\% & 1.6\% & 1.8\%\\
				infotocap & \textbf{83.4\%} & 4.9\% & 2.0\% & 2.2\% & 3.1\% & 2.2\% & 2.1\%\\
				mp42aac & \textbf{73.0\%} & 4.0\% & 3.4\% & 4.5\% & 5.1\% & 4.9\% & 5.1\%\\
				mp4tag & \textbf{78.9\%} & 4.3\% & 3.5\% & 3.1\% & 4.2\% & 3.5\% & 2.5\%\\
				nm & \textbf{77.4\%} & 4.7\% & 3.7\% & 5.4\% & 3.3\% & 2.3\% & 3.2\%\\
				pdfinfo & \textbf{81.1\%} & 4.3\% & 3.1\% & 3.2\% & 3.2\% & 2.6\% & 2.6\%\\
				readelf & \textbf{81.5\%} & 5.1\% & 2.8\% & 2.3\% & 2.1\% & 3.8\% & 2.3\%\\
				tiff2pdf & \textbf{68.3\%} & 3.7\% & 4.3\% & 6.4\% & 6.2\% & 5.8\% & 5.3\%\\
				tiff2ps & \textbf{83.1\%} & 4.4\% & 2.8\% & 2.0\% & 2.5\% & 2.8\% & 2.5\%\\
				txtext & \textbf{83.5\%} & 4.4\% & 1.8\% & 3.5\% & 2.0\% & 2.8\% & 2.0\%\\
				xmllint & \textbf{84.7\%} & 4.8\% & 2.8\% & 2.3\% & 2.2\% & 1.5\% & 1.7\%\\
				\toprule
				Average & \textbf{78.7\%} & 4.4\% & 3.2\% & 3.6\% & 3.5\% & 3.5\% & 3.1\%\\
				\bottomrule
			\end{tabular}
		}
	\end{table}

2) Table \ref{table_com_sel} and Figure \ref{fig_com_sel} (in Appendix) show the distribution of the selected objectives. As we take 3 objectives 
(speed, stack, and cmp) into consideration, there are 8 objective combinations in total. The most significant observation of Table \ref{table_com_sel} and Figure \ref{fig_com_sel} is that the MPMAB model tends to select the speed objective in more than 60\% of all the combination selections. Again, we clarify our agreement with previous work \cite{zeror, fullspeed, instrcr}: the top priority of fuzzing is execution speed. Therefore, in Equation \ref{equ_reward}, we add a penalty to objectives that slow down the fuzzing process. It is in line with our tendency toward speed and can also explain the features of Table \ref{table_com_sel} and Figure \ref{fig_com_sel}.

	\begin{figure}[htb]
		\centering
		\includegraphics[width=0.9\columnwidth]{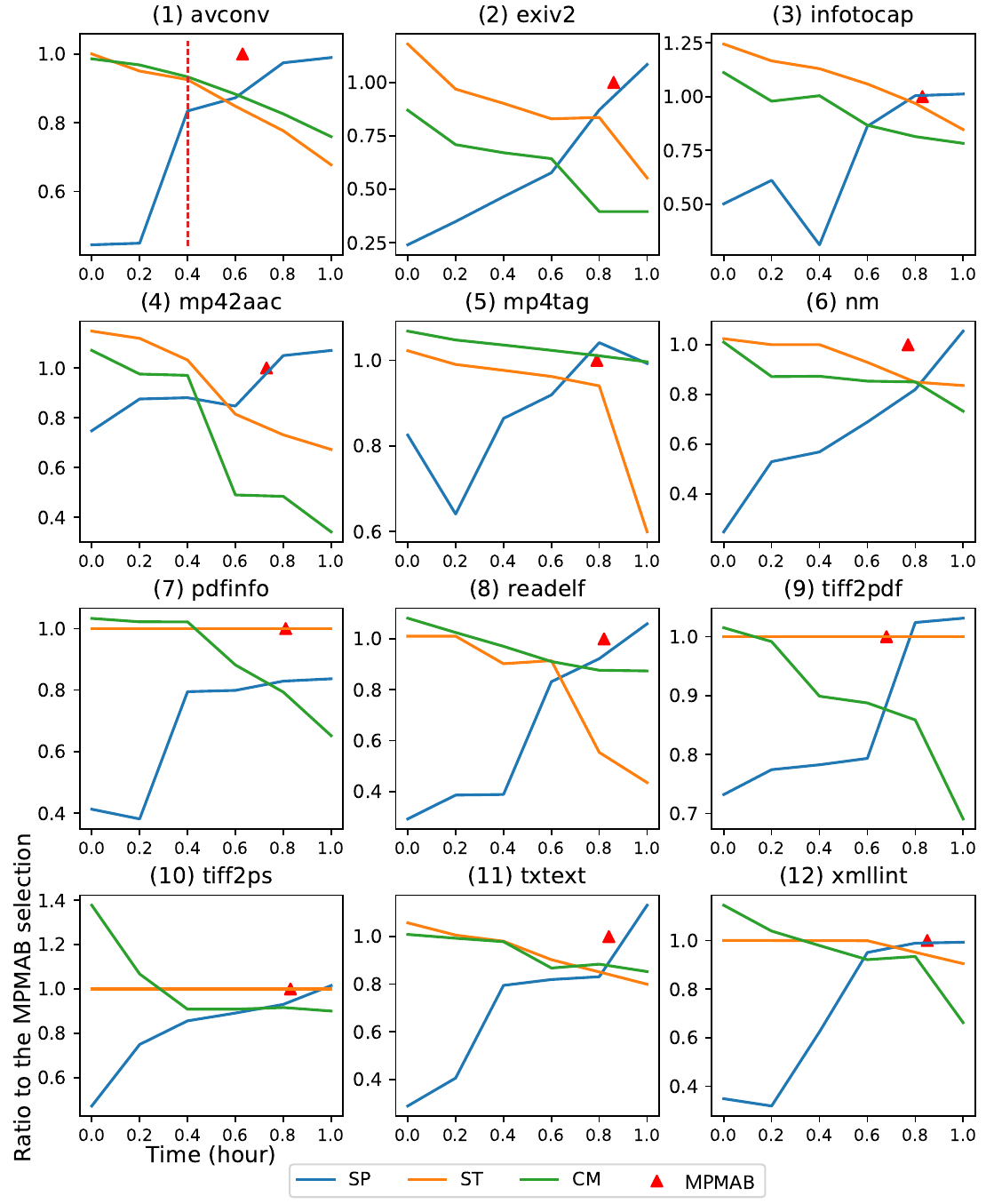}
		\caption{\footnotesize{The $\mathtt{X}$-axis denotes different selection strategies. The $\mathtt{Y}$-axis is the ratio of objective values ($\mathtt{\frac{v'}{v_{\scriptscriptstyle M}}}$) in different selection strategies ($\mathtt{v'}$) in contrast to our MPMAB selection ($\mathtt{v_{\scriptscriptstyle M}}$). $\mathtt{\frac{v'}{v_{\scriptscriptstyle M}}} < \mathtt{1.0}$ means the objective value of this selection is less than the MPMAB selection and vice versa.}}
		\label{fig_com_diff}
	\end{figure}

\begin{table*}
	\setlength{\abovecaptionskip}{0cm}
	\caption{Average energy consumption to reach the objective values of the fuzzers}
	\label{table_energy}
	\centering
	\resizebox{0.9\textwidth}{!}{
		\begin{threeparttable}
			\begin{tabular}{ccccccccccccc}
				\bottomrule
				\multirow{2}*{\textbf{Targets}} & \multicolumn{4}{c}{\textbf{Ave. energy of execution speed}} & \multicolumn{4}{c}{\textbf{Ave. energy of stack memory}} & \multicolumn{4}{c}{\textbf{Ave. energy of satisfied comparison bytes}} \\
				\cline{2-13}
				\bfseries ~ & \bfseries MobFuzz & \textbf{AFL} & \textbf{MemLock} & \textbf{FuzzFactory} & \bfseries MobFuzz & \textbf{AFL} & \textbf{MemLock} & \textbf{FuzzFactory} & \bfseries MobFuzz & \textbf{AFL} & \textbf{MemLock} & \textbf{FuzzFactory} \\
				\toprule
				avconv & \textbf{29.33}\tnote{1} & 39.28 & 39.06 & 40.33 & \textbf{83.45} & 90.98 & 140.22 & 128.97 & \textbf{0.01} & 0.03 & 0.03 & 0.02\\
				exiv2 & \textbf{22.03} & 40.73 & 37.59 & 32.22& \textbf{10.80} & 19.04 & 16.23 & 11.01 & \textbf{0.16} & 0.37 & 0.47 & 0.29\\
				infotocap & \textbf{19.27} & 25.51 & 30.14 & 30.01 & \textbf{368.35} & 548.58 & 376.58 & 502.18 & \textbf{0.03} & 0.08 & 0.08 & 0.05 \\
				mp42aac & \textbf{32.97} & 33.32 & 33.79 & 34.311 & \textbf{173.50} & 455.01 & 383.46 & 420.23 & \textbf{0.96} & 2.75 & 1.77 & 1.86 \\
				mp4tag & \textbf{29.25} & 32.90 & 33.24 & 33.428 & \textbf{248.16} & 322.99 & 287.42 & 358.06 & \textbf{0.75} & 1.61 & 1.76 & 1.25\\
				nm & \textbf{25.78} & 33.18 & 32.04 & 32.79 & 216.21 & 449.31 & 277.64 & \textbf{34.86} & \textbf{0.31} & 0.66 & 0.66 & 0.44\\
				pdfinfo & \textbf{31.19} & 33.59 & 33.77 & 32.91 & 1.04 & \textbf{0.73} & 0.81 & 0.73 & \textbf{1.46} & 2.02 & 3.26 & 2.72\\
				readelf & \textbf{29.88} & 33.66 & 32.61 & 33.82 & \textbf{8765.65} & 19732.93 & 9243.49 & 11510.45 & \textbf{0.13} & 0.81 & 0.69 & 0.47\\
				tiff2pdf & \textbf{28.95} & 33.83 & 33.15 & 31.88 & \textbf{958.27} & 2346.60 & 1117.60 & 2712.60 & \textbf{1.84} & 1.94 & 2.23 & 1.87\\
				tiff2ps & \textbf{30.16} & 33.27 & 33.99 & 32.35 & \textbf{4.03} & 7.38 & 7.47 & 7.39 & \textbf{3.63} & 4.75 & 6.01 & 5.90\\
				txtext & \textbf{29.06} & 32.70 & 33.98 & 34.39 & \textbf{0.41} & 0.55 & 0.80 & 0.66 & \textbf{1.22} & 1.86 & 3.05 & 2.96\\
				xmllint & \textbf{29.20} & 32.64 & 31.77 & 33.05 & \textbf{24.39} & 163.85 & 140.92 & 260.80 & \textbf{0.12} & 0.46 & 0.23 & 0.41\\
				\toprule
				Average & \textbf{28.1} & 33.7(\footnotesize{-17\%})\tnote{2} & 33.8(\footnotesize{-17\%}) & 33.5(\footnotesize{-16\%}) & \textbf{904.5} & 2011.5(\footnotesize{-55\%}) & 999.4(\footnotesize{-10\%}) & 1329.0(\footnotesize{-32\%}) & \textbf{0.87} & 1.44(\footnotesize{-39\%}) & 1.68(\footnotesize{-48\%}) & 1.52(\footnotesize{-42\%})\\
				\bottomrule
			\end{tabular}
			\begin{tablenotes}
				\footnotesize
				\item[1] Smaller values are better. $^2$ The percentages in the brackets of the last line denote the decease in contrast to the baseline fuzzers.
			\end{tablenotes}
		\end{threeparttable}
	}
\end{table*}

3) Finally, we need to prove whether we select the best objective combination. Figure \ref{fig_com_diff} shows different strategies in contrast to ours. The $\mathtt{X}$-axis is the proportion of speed in all the selected combinations, and different proportions indicate different selection strategies. We set 6 strategies, including 0\%, 20\%, 40\%, 60\%, 80\%, and 100\% proportions of speed. The $\mathtt{Y}$-axis shows the ratio of objective values ($\mathtt{\frac{v'}{v_{\scriptscriptstyle M}}}$) of different strategies ($\mathtt{v'}$) in contrast to our MPMAB selection ($\mathtt{v_{\scriptscriptstyle M}}$). $\mathtt{\frac{v'}{v_{\scriptscriptstyle M}}}$ less than $\mathtt{1.0}$ means the objective value of this selection is less than the MPMAB selection, and this selection is worse. For example, in $\mathtt{avconv}$, when the proportion of speed is 40\%, we highlight the results with a red dotted line. The speed of this selection strategy is approximately 80\% of ours. The stack and cmp are just above 90\% of ours.

As the proportion of speed increases, the value of speed also increases. However, as discussed previously in this paper, it is often the case that objectives have opposite effects on each other. The values of stack and cmp decrease as speed increases. More interestingly, we mark the result of MPMAB selection in the figure as a red triangle. The red triangle is close to the intersection of the three lines. In conclusion, none of the 6 selection strategies outperforms the MPMAB selection. The objective values of our strategy are all greater than those of the 6 strategies. The figure indicates that our combination selection method handles the relationship among the multiple objectives, and we can choose the most appropriate proportion of each combination to optimize all the objectives. According to our discussion above, we can answer RQ2.

$\bullet$ \textit{Answer to RQ2: Our selection strategy can adaptively select the best objective combinations.}


\subsection{Power schedule}

	\begin{figure}[htb]
		\centering
		\includegraphics[width=0.7\columnwidth]{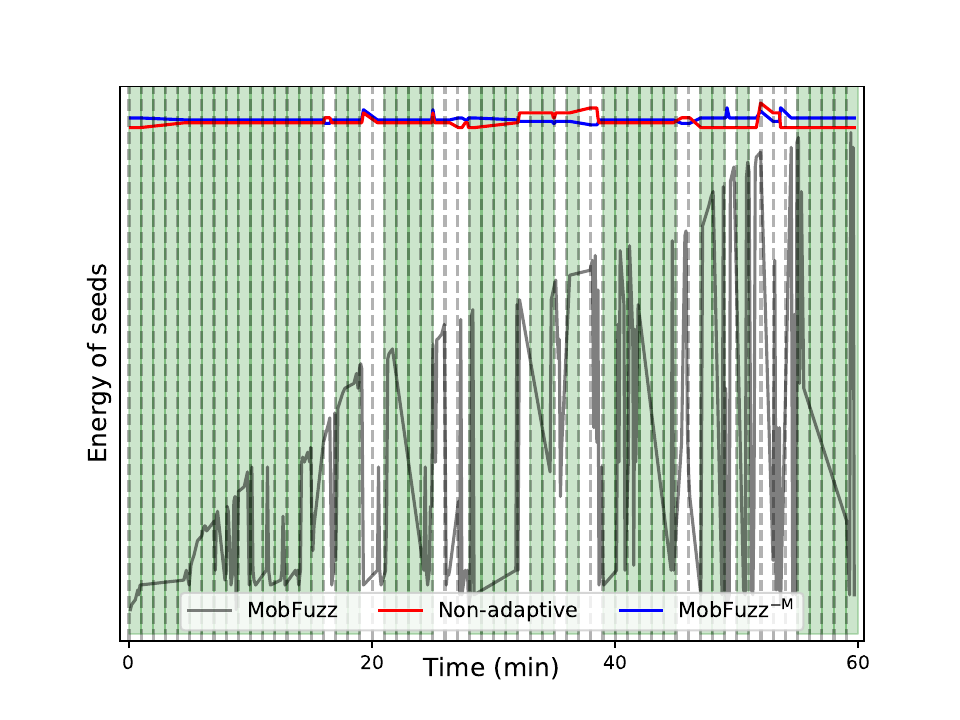}
		\caption{\footnotesize{Comparison of our adaptive power schedule with non-adaptive and MobFuzz$^{\scriptscriptstyle \rm{-M}}$ (MobFuzz without MPMAB) schedules in 1 hour. The $\mathtt{Y}$-axis shows the amount of energy allocated by the schedules. The green background color denotes that in this time interval, our adaptive schedule adjusts the energy according to the chosen objective combination and the non-adaptive schedule fails to.}}
		\label{fig_pow_diff}
	\end{figure}

As described above, there is no previous work solving the problem of energy allocation under the chosen objective combination. To demonstrate that our adaptive power schedule can allocate a different amount of energy to seeds under the chosen objective combination, we show the comparison among our schedule, the non-adaptive (i.e., the power schedule of AFL) schedule, and MobFuzz$^{\scriptscriptstyle \rm{-M}}$ (MobFuzz without MPMAB) schedule in Figure \ref{fig_pow_diff}. The $\mathtt{X}$-axis shows one hour of the fuzzing process. The $\mathtt{Y}$-axis shows the energy allocated by our adaptive schedule and others. Moreover, the $\mathtt{X}$-axis is divided into 60 minutes, and each minute represents a time interval with different chosen objective combinations. Our schedule adaptively allocates energy according to the chosen combination, where the non-adaptive schedule is insensitive to the changes and keeps allocating the same amount of energy under different objective combinations. Our schedule outperforms the non-adaptive schedule in 43 out of the 60 minutes, which demonstrates the effectiveness of our adaptive power schedule.

In addition, the result of MobFuzz$^{\scriptscriptstyle \rm{-M}}$ schedule is similar to the non-adaptive schedule. This result indicates that without MPMAB, MobFuzz cannot adaptively choose the best objective combination or allocate the appropriate amount of energy. Therefore, there is no change in the allocated energy corresponding to the chosen objective combination. In conclusion, the comparison between MobFuzz and MobFuzz$^{\scriptscriptstyle \rm{-M}}$ proves the effectiveness of MPMAB.

To answer whether our power schedule saves energy, we conduct the following experiments. Table \ref{table_energy} shows the average energy consumption to reach the objective values in Table \ref{table_objective}. We divide it by the number of executions for these objective values and calculate the average energy. As we can see from the table, among all the 108 pairs of comparisons with the baseline fuzzers, only in 4 of them does our power schedule have greater average energy consumption, and these are in $\mathtt{nm}$ and $\mathtt{pdfinfo}$. In other words, MobFuzz allocates less energy for the objectives in over 96\% of the scenarios. For instance, in $\mathtt{readelf}$, our allocated energy is 6x less (0.13 vs. 0.81) than that of AFL. In the $\mathtt{Average}$ row of the table, we calculate the averages of all the values. Among all the average values, MobFuzz allocates less energy. Specifically, in comparison with AFL in terms of stack memory, we save more than 50\% of energy. From the above discussion, we can answer RQ3.

$\bullet$ \textit{Answer to RQ3: Our power schedule can adaptively allocate energy according to the chosen objective combination and save more energy compared with the baseline fuzzers.}

\subsection{Evaluation on The NIC Algorithm}

\subsubsection{Results of Good Seeds}

	\begin{figure}[htb]
		\centering
		\includegraphics[width=0.9\columnwidth]{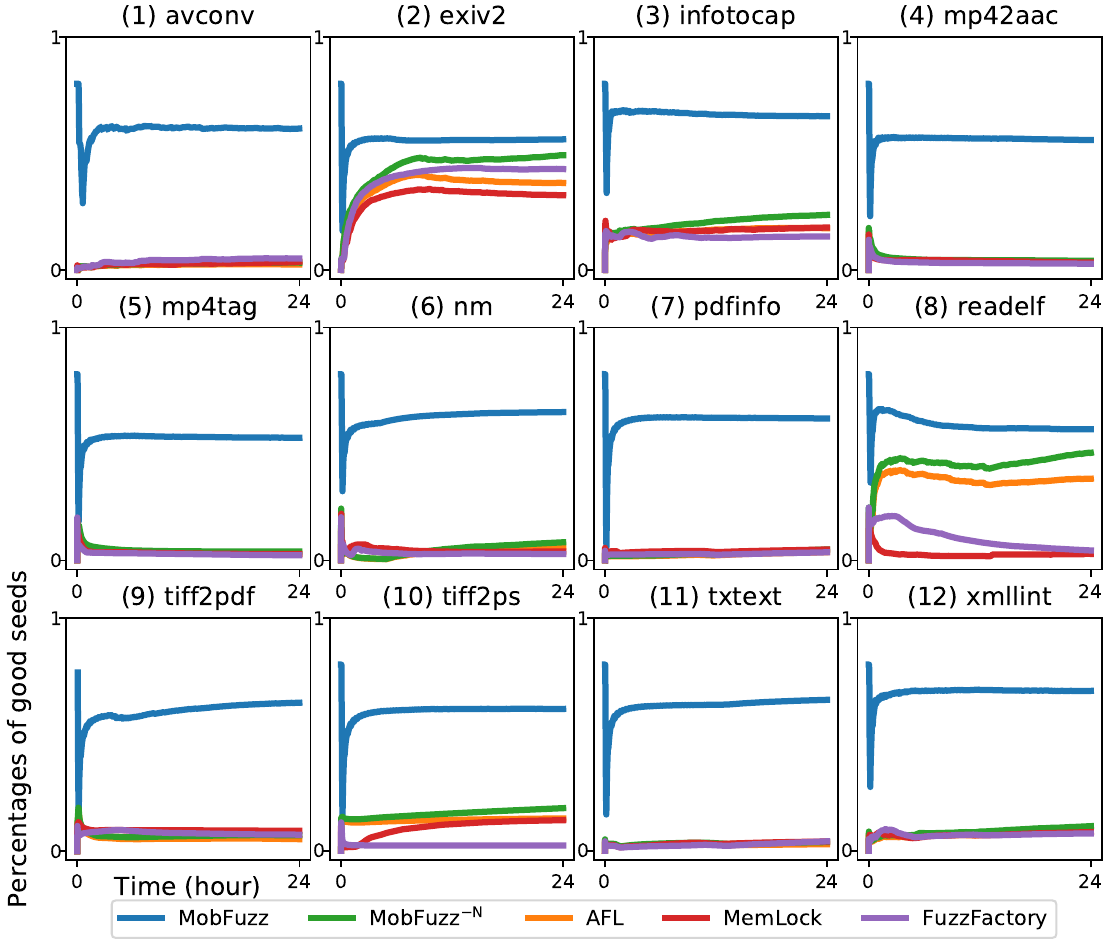}
		\caption{\footnotesize{Percentages of good seeds (seeds that achieve greater objective values than the average in the selected objective combination) generated during fuzzing, where greater values are better. MobFuzz$^{\scriptscriptstyle \rm{-N}}$ denotes MobFuzz without NIC.}}
		\label{fig_good_seed}
	\end{figure}

\begin{table*}
	\setlength{\abovecaptionskip}{0cm}
	\caption{Comparison of the main loop mutation operators (Main) and NIC mutation operators}
	\label{table_new_muter}
	\centering
	\resizebox{0.8\textwidth}{!}{
		\begin{threeparttable}
			\begin{tabular}{ccccccccccc}
				\bottomrule
				\multirow{2}*{\textbf{Targets}} & \multicolumn{2}{c}{\textbf{Ave. Time (ms)}} & \multicolumn{2}{c}{\textbf{Ave. Length (byte)}} & \multicolumn{2}{c}{\textbf{Num. of executions}} & \multicolumn{2}{c}{\textbf{Num. of seeds}} & \multicolumn{2}{c}{\textbf{Executions per seed}}\\
				\cline{2-11}
				\bfseries ~ & \textbf{NIC} & \textbf{Main} & \textbf{NIC} & \textbf{Main} & \textbf{NIC} & \textbf{Main} & \textbf{NIC} & \textbf{Main} & \textbf{NIC} & \textbf{Main}\\
				\toprule
				avconv & \textbf{10.27} & 10.65 & \textbf{1273.19} & 5886.82 & 2.00$*10^6$ & 5.38$*10^6$ & 1796.20 & 20367.00 & 1113.46 & \textbf{264.15}\\
				exiv2 & \textbf{1.13} & 1.26 & \textbf{259.16} & 4140.21 & 3.24$*10^5$ & 5.76$*10^7$ & 74.40 & 5960.80 & \textbf{4354.83} & 9663.13\\
				infotocap & \textbf{1.58} & 1.94 & \textbf{8164.61} & 17377.55 & 3.20$*10^5$ & 3.31$*10^7$ & 195.60 & 5490.30 & \textbf{1635.99} & 6028.81\\
				mp42aac & \textbf{0.45} & 0.52 & \textbf{1201.41} & 5421.39 & 88873.40 & 5.31$*10^8$ & 53.70 & 3192.90 & \textbf{1654.99} & 16630.65\\
				mp4tag & \textbf{0.44} & 0.64 & \textbf{1230.61} & 5570.29 & 95440.40 & 1.24$*10^8$ & 39.80 & 3061.10 & \textbf{2398.00} & 40508.31\\
				nm & \textbf{0.59} & 0.62 & \textbf{1743.88} & 5450.27 & 95241.40 & 1.28$*10^8$ & 37.60 & 3256.10 & \textbf{2533.01} & 39310.83\\
				pdfinfo & \textbf{1.06} & 1.09 & \textbf{2605.15} & 2652.75 & 499.10 & 7.57$*10^7$ & 0.40 & 40.30 & \textbf{1247.75} & 1.88$*10^6$\\
				readelf & \textbf{0.44} & 0.54 & \textbf{1039.05} & 9445.07 & 1.60$*10^7$ & 1.02$*10^8$ & 7671.90 & 49200.00 & 2085.53 & \textbf{2073.17}\\
				tiff2pdf & \textbf{0.45} & 0.50 & \textbf{759.55} & 4170.77 & 14367.80 & 1.64$*10^8$ & 12.60 & 1213.00 & \textbf{1140.30} & 1.35$*10^5$\\
				tiff2ps & \textbf{0.45} & 0.46 & \textbf{372.43} & 4924.48 & 1472.60 & 1.77$*10^8$ & 4.00 & 398.50 & \textbf{368.15} &4.44$*10^5$\\
				txtext & \textbf{1.20} & 1.21 & \textbf{3289.64} & 4360.13 & 398.00 & 6.85$*10^7$ & 5.20 & 41.00 & \textbf{76.53} & 1.67$*10^6$ \\
				xmllint & \textbf{0.75} & 0.77 & \textbf{3629.15} & 9838.51 & 3.92$*10^5$ & 9.02$*10^7$ & 594.50 & 6683.40 & \textbf{658.38} & 13496.12\\
				\toprule
				Average & \textbf{1.45} & 1.54(\footnotesize{-5.5\%})\tnote{1} & \textbf{1527.7} & 2808.6(\footnotesize{-45.6\%}) & 2.46$*10^6$ & 4.10$*10^7$(\footnotesize{P5.7\%})\tnote{2} & 12867.2 & 88471.6(\footnotesize{P12.7\%}) & \textbf{1914.2} & 4632.6(\footnotesize{-58.7\%})\\
				\bottomrule
			\end{tabular}
			\begin{tablenotes}
				\footnotesize
				\item[1] The percentages with a “-” in the brackets of the last line denote the improvements in contrast to Main. $^2$ The percentages with a “P” denote the proportion of NIC to the total (Main+NIC).
			\end{tablenotes}
		\end{threeparttable}
	}
\end{table*}

We define seeds that achieve greater objective values than the average in the selected objective combination as \textbf{good seed}s. Figure \ref{fig_good_seed} shows the percentages of good seeds generated in MobFuzz and the baseline fuzzers. We can extract two conclusions from the figure. First, among all the 12 target programs, the percentages of good seeds in MobFuzz are greater than those of the baseline fuzzers. The minimum performance improvement is in $\mathtt{readelf}$, which is approximately {2x}. As discussed above, in NIC, we optimize the values with crossover, mutation, and execution. The results show that the NIC algorithm in MobFuzz can produce more seeds that are better than the average level, which helps optimize the objectives. Second, without the help of the NIC algorithm, the performance of the baseline fuzzers is similar, which has small percentages of good seeds.

Moreover, we compare MobFuzz with MobFuzz$^{\scriptscriptstyle \rm{-N}}$ (MobFuzz without NIC) in Figure \ref{fig_good_seed}. When NIC is disabled, we can see a noticeable difference between MobFuzz and MobFuzz$^{\scriptscriptstyle \rm{-N}}$. The percentage of good seeds in MobFuzz$^{\scriptscriptstyle \rm{-N}}$ decreases to the level of the baseline fuzzers without NIC. This result indicates that without NIC, MobFuzz$^{\scriptscriptstyle \rm{-N}}$ cannot generate as many good seeds with greater objective values as MobFuzz. In conclusion, by comparing MobFuzz with MobFuzz$^{\scriptscriptstyle \rm{-N}}$, we demonstrate the effectiveness of NIC.

Along with the results in Figure \ref{fig_pow_diff}, we can conclude that there is synergy between MPMAB and NIC: 1) The power schedule and NIC are executed under the chosen combination. 2) NIC outputs the Pareto seeds with the optimal values. These values can affect the combination selection and power schedule in return. In this way, these parts can cooperate. The combination of MPMAB and NIC generates the best result.

\subsubsection{Results of the Mutation Operators}

Table \ref{table_new_muter} is the comparison between the mutation operator of NIC and the main fuzzing loop (i.e., the operators of AFL). We can also draw two conclusions based on the table. First, according to the table, in the $\mathtt{Time}$ and $\mathtt{Length}$ columns, NIC has less time consumption and shorter length in all of the 24 pairs of comparisons. For the $\mathtt{tiff2ps}$ target program, NIC is 13x less than Main. Regarding the average values of time and length, NIC also outperforms Main. There is 5.5\% less time and 45.6\% less length compared with Main. The results demonstrate that with the help of our mutation strategy of selecting operators with better performance according to the chosen objective combination, NIC can achieve better objective optimization.

Second, we can determine the path discovery efficiency of NIC based on the $\mathtt{Executions}$ and $\mathtt{Seeds}$ columns. The numbers of executions of Main (i.e., the main loop) and NIC show how many seeds are mutated and executed in each part. The executions of NIC are 5.7\% of the total (Main+NIC). If we disable the shared seed pool mechanism, the inputs that cover new paths in NIC will not be saved in the seed pool, and this 5.7\% of the executions will be wasted. Moreover, NIC also finds seeds with higher efficiency. NIC generates 12.7\% of the total seeds with only 5.7\% of the executions. The last two columns show the executions per seed of Main and NIC. These values indicate the number of executions required to find a new seed. We can see a 58.7\% decrease in the comparison, which also proves that NIC can generate new seeds more efficiently.


\subsubsection{Performance Overhead of NIC}

	\begin{table}[htb]
		\setlength{\abovecaptionskip}{0cm}
		\caption{Performance overhead of the NIC algorithm and performance improvement of our adopted techniques in NIC}
		\label{table_nic_speed}
		\centering
		\resizebox{0.8\columnwidth}{!}{
			\begin{threeparttable}
			\begin{tabular}{ccccc}
				\bottomrule
				\bfseries  & \textbf{MobFuzz} & \tnote{1} \textbf{MobFuzz$^{\scriptscriptstyle \rm{-N}}$} & \tnote{2} \textbf{MobFuzz$\scriptscriptstyle \rm ^{-POR}$} \\
				\toprule
				Speed & \textbf{1147.51} & 1186.52(-3.3\%) & 1004.07(+14.3\%)\\
				\bottomrule
			\end{tabular}
			\begin{tablenotes}
				\footnotesize
				\item[1] MobFuzz$\scriptscriptstyle \rm ^{-N}$ denotes MobFuzz without NIC.
				\item[2] MobFuzz$\scriptscriptstyle \rm ^{-POR}$ denotes MobFuzz with NIC but without the performance overhead reduction.
			\end{tablenotes}
			\end{threeparttable}
		}
	\end{table}

Table \ref{table_nic_speed} shows the average execution speed of MobFuzz and two other configurations. MobFuzz$\scriptscriptstyle \rm ^{-N}$ denotes MobFuzz without the NIC algorithm, and MobFuzz$\scriptscriptstyle \rm ^{-POR}$ denotes MobFuzz with NIC but without the performance overhead reduction in NIC. First, enabling the NIC algorithm in MobFuzz brings about 3.3\% performance overhead to the fuzzing process, which is acceptable. Though NIC slightly slows down the fuzzing process, it can produce the optimal results for multi-objectives. By sacrificing this 3.3\% of speed, we can get optimal results for other objectives. Additionally, even with this 3.3\% performance overhead, MobFuzz is still faster than the baseline fuzzers, which is shown in Table \ref{table_objective}. Second, the comparison with MobFuzz$\scriptscriptstyle \rm ^{-POR}$ indicates the performance improvement of the techniques in NIC. The shared seed pool and other techniques increase the speed of fuzzing by 14.3\%, which demonstrates the effectiveness of our techniques to reduce the performance overhead in NIC.


\subsubsection{Comparison with Black-box MOO Techniques}

	\begin{table}[htb]
		\setlength{\abovecaptionskip}{0cm}
		\caption{Comparison with NSGA-II in the objective values}
		\label{table_nic_blackbox}
		\centering
		\resizebox{0.9\columnwidth}{!}{
			\begin{threeparttable}
			\begin{tabular}{ccccc}
				\bottomrule
				\bfseries  & \textbf{Execution speed} & \textbf{Stack memory} & \textbf{Satisfied comparison bytes} \\
				\toprule
				MobFuzz & {1147.51} & {8.72}${*10^5}$ & {9.60}${*10^7}$ \\
				\tnote{1} MobFuzz$\scriptscriptstyle \rm ^{N2}$ & 327.20 & 1.48${*10^5}$ & 8.83${*10^6}$ \\
				\tnote{2} MobFuzz$\scriptscriptstyle \rm ^{N2+APS}$ & 559.31 & 2.97${*10^5}$ & 1.09${*10^7}$\\
				\tnote{3} MobFuzz$\scriptscriptstyle \rm ^{N2+CMO}$ & 446.89 & 6.87${*10^5}$ & 7.63${*10^7}$\\
				\tnote{4} MobFuzz$\scriptscriptstyle \rm ^{N2+POR}$ & 986.37 & 3.02${*10^5}$ & 2.56${*10^7}$\\
				\bottomrule
			\end{tabular}
			\begin{tablenotes}
				\footnotesize
				\item[1] MobFuzz$\scriptscriptstyle \rm ^{N2}$ denotes replacing NIC with NSGA-II.
				\item[2] MobFuzz$\scriptscriptstyle \rm ^{N2+APS}$ denotes enabling adaptive population size in MobFuzz$\scriptscriptstyle \rm ^{N2}$.
				\item[3] MobFuzz$\scriptscriptstyle \rm ^{N2+CMO}$ denotes enabling co-mutation operators in MobFuzz$\scriptscriptstyle \rm ^{N2}$.
				\item[4] MobFuzz$\scriptscriptstyle \rm ^{N2+POR}$ denotes enabling performance overhead reduction in MobFuzz$\scriptscriptstyle \rm ^{N2}$.
			\end{tablenotes}
		\end{threeparttable}
		}
	\end{table}

To compare with existing black-box MOO techniques such as NSGA-II \cite{nsga}, we replace key design components in MobFuzz with applicable existing designs from NSGA-II and call it MobFuzz$\scriptscriptstyle \rm ^{N2}$. Table \ref{table_nic_blackbox} shows the comparison of MobFuzz with MobFuzz$\scriptscriptstyle \rm ^{N2}$ in the objective values. In Section \ref{sec_nic}, we introduce three aspects of techniques in NIC, including adaptive population size, co-mutation operators, and overhead reduction. None of these techniques is adopted in MobFuzz$\scriptscriptstyle \rm ^{N2}$.

MobFuzz$\scriptscriptstyle \rm ^{N2}$ uses a fixed initial population size. At the beginning of the fuzzing campaign, this size is too large for the small seed pool. Starting with this large population will slow down the fuzzing process and contribute nothing to increase the objective values. When the seed pool becomes large, this fixed initial population size is not enough. It lacks diversity and cannot produce the optimal result. Therefore, we use adaptive population size to handle these issues. Moreover, by enabling the adaptive population size in MobFuzz$\scriptscriptstyle \rm ^{N2}$, we use the result of MobFuzz$\scriptscriptstyle \rm ^{N2+APS}$ to demonstrate the effectiveness of this technique. MobFuzz$\scriptscriptstyle \rm ^{N2+APS}$ achieves greater values than MobFuzz$\scriptscriptstyle \rm ^{N2}$ in the three objectives.

Co-mutation operators are introduced in NIC. The original mutation of NSGA-II does not apply to a fuzzing situation. We remove the ineffective mutation operators in NSGA-II. Moreover, we build connections between objectives and mutation operators. The best operators are selected according to the objectives. In Table \ref{table_nic_blackbox}, the comparison of MobFuzz$\scriptscriptstyle \rm ^{N2}$ with MobFuzz$\scriptscriptstyle \rm ^{N2+CMO}$ demonstrates the effectiveness of the co-mutation operators. All the objective values are greater in MobFuzz$\scriptscriptstyle \rm ^{N2+CMO}$.

In MobFuzz$\scriptscriptstyle \rm ^{N2}$, the seeds generated in NSGA-II are independent of the seeds in the main fuzzing loop. Only at the end of the evolutionary process, a small number of seeds are saved to the main seed pool. This independence of the main fuzzing loop wastes the executions during NSGA-II, and it is the primary performance overhead in MobFuzz$\scriptscriptstyle \rm ^{N2}$. We propose a shared seed pool technique in MobFuzz to handle this issue. This technique connects the NIC to the main fuzzing loop and saves seeds during the evolutionary process of NIC. In addition, by enabling the performance overhead reduction in MobFuzz$\scriptscriptstyle \rm ^{N2}$, the speed result of MobFuzz$\scriptscriptstyle \rm ^{N2+POR}$ is much greater than MobFuzz$\scriptscriptstyle \rm ^{N2}$. This demonstrates the effectiveness of our overhead reduction techniques.

$\bullet$ \textit{Answer to RQ4: NIC can optimize the objectives without introducing additional performance overhead.}

\subsection{MAGMA Data Set}

MAGMA \cite{hazimeh2020magma} is a newly proposed data set. It contains 7 projects with 19 target programs. MAGMA is a ground-truth fuzzing benchmark based on real programs with real bugs, which allows for a fair and accurate evaluation of fuzzers.

\textbf{Baseline fuzzers to compare. }Following the experiments in the MAGMA paper, AFL\cite{afl}, AFLFast\cite{bohme2017coverage}, AFL++\cite{afl++} (with $\mathtt{lto}$ and $\mathtt{cmplog}$ enabled), FairFuzz\cite{fairfuzz}, honggfuzz\cite{honggfuzz}, MOPT \cite{mopt}, and SYMCC \cite{poeplau2020symbolic} are used in our evaluation. The AFL-based fuzzers use consistent settings: deterministic and havoc. Using this configuration can comprehensively test the vulnerability detection ability of MobFuzz compared with state-of-the-art fuzzers.

We use seeds in the $\mathtt{corpus}$ directory provided by MAGMA as the initial seeds. Our evaluations are conducted on three servers for 10 times and 24 hours.

Experiments in this subsection also follow the configuration in the MAGMA paper, which are divided into two parts: 1) The number of bugs discovered by the fuzzers (the $\mathtt{unique\_bugs}$ results counted by the MAGMA tool script). Table \ref{table_magma_bug} shows the results of bugs, and Table \ref{table_magma_bug_p} shows the p values of the results. 2) Time to bug (the $\mathtt{TTB}$ results). Table \ref{table_magma_ttb} shows the TTB results, and Table \ref{table_magma_ttb_p} shows the p values of the results. Additionally, the TTB results of each bug are in Table \ref{table_magma_ttb_detail} and \ref{table_magma_ttb_detail2} in Appendix.


\subsubsection{Bug Count}

\begin{table}[h]
	\setlength{\abovecaptionskip}{0cm}
	\caption{Average number of MAGMA bugs found by the fuzzers}
	\label{table_magma_bug}
	\centering
	\resizebox{0.95\columnwidth}{!}{
		\begin{tabular}{ccccccccc}
			\bottomrule
			\scriptsize{\textbf{Targets}} & \scriptsize{\textbf{AFL}} & \scriptsize{\textbf{AFLFast}} & \scriptsize{\textbf{AFL++}} & \scriptsize{\textbf{FairFuzz}} & \scriptsize{\textbf{honggfuzz}} & \scriptsize{\textbf{MOPT}} & \scriptsize{\textbf{SYMCC}} & \scriptsize{\textbf{MobFuzz}} \\
			\hline 
			libpng & 4.1  & 4.3  & 4.2  & 3.9  & 4.3  & 4.0  & 4.2 & 4.3 \\
			libtiff & 8.3  & 8.0  & 8.2  & 6.9  & 6.0  & 7.5 & 6.7 & \textbf{8.5}\\
			libxml2 & 6.0  & 6.1  & 6.0  & 6.0  & 6.5  & 6.8 & 5.5 & \textbf{7.0}\\
			openssl & 3.8  & 3.8  & {6.5}  & 4.1  & 6.1 & 6.0 & 6.5 & 6.4\\
			php & 6.0  & 6.5  & 5.9  & 5.7  & 6.2  & 6.5 & 5.5 & 6.2\\
			poppler & 7.0  & 7.0  & 7.2  & 7.0  & 6.5  & 7.3 & 6.4 & \textbf{10.9}\\
			sqlite3 & 3.8  & 5.0  & 2.7  & 5.0  & 2.7  & 1.9 & 2.0 & \textbf{5.1}\\
			\toprule
			Average & 5.6  & 5.8  & 5.8  & 5.5  & 5.5  & 5.7 & 5.3 & \textbf{6.9}\\
			\bottomrule
		\end{tabular}
	}
\end{table}

Table \ref{table_magma_bug} shows the bugs discovered by the fuzzers, and Table \ref{table_magma_bug_p} shows the p values of these results. According to the results, MobFuzz outperforms all the baseline fuzzers in 4 out of the 7 projects. The average number of bugs of MobFuzz is greater than other fuzzers. Additionally, in $\mathtt{sqlite3}$, MobFuzz has the maximum improvement against the baseline fuzzers, which is about 3x better than MOPT. One advantage of MobFuzz over AFL, AFLFast, FairFuzz, honggfuzz, and MOPT is the satisfied comparison bytes MobFuzz concentrates on. By satisfying more bytes, some of the MAGMA bugs can be triggered more easily in MobFuzz. For instance, Bug SQL003 in Table \ref{table_magma_ttb_detail2} is only triggered when $\mathtt{if (!data)}$ is satisfied. MobFuzz succeeds in 9 hours, and others fail to.


\begin{table}[h]
	\setlength{\abovecaptionskip}{0cm}
	\caption{p values of the results in Table \ref{table_magma_bug}}
	\label{table_magma_bug_p}
	\centering
	\resizebox{0.95\columnwidth}{!}{
		\begin{tabular}{ccccccccc }
			\bottomrule
			\scriptsize{\textbf{Targets}} & \scriptsize{\textbf{AFL}} & \scriptsize{\textbf{AFLFast}} & \scriptsize{\textbf{AFL++}} & \scriptsize{\textbf{FairFuzz}} & \scriptsize{\textbf{honggfuzz}} & \scriptsize{\textbf{MOPT}} & \scriptsize{\textbf{SYMCC}} & \scriptsize{\textbf{MobFuzz}} \\
			\hline 
			libpng & $<10^{-4}$  & 0.04  & $<10^{-3}$  & $<10^{-3}$  & 0.01  & 0.02 & 0.05 & $<10^{-3}$ \\
			libtiff & $<10^{-5}$  & $<10^{-3}$  & 0.02  & $<10^{-4}$  & 0.06  & 0.45 & $<10^{-4}$  & 0.01\\
			libxml2 & $<10^{-5}$  & $<10^{-4}$  & $<10^{-3}$  & 0.06  & $<10^{-4}$  & $<10^{-3}$ & 0.03 & $<10^{-4}$\\
			openssl & 0.03  & $<10^{-4}$  & 0.08  & $<10^{-3}$  & 0.01  & $<10^{-3}$ & $<10^{-3}$ & $<10^{-4}$\\
			php & $<10^{-3}$  & 0.25  & $<10^{-4}$  & 0.05  & 0.02 & $<10^{-3}$ & 0.22  & 0.02\\
			poppler & $<10^{-4}$  & $<10^{-3}$  & 0.02  & 0.07  & $<10^{-4}$  & 0.03 & 0.01  & $<10^{-5}$\\
			sqlite3 & 0.07  & $<10^{-3}$  & 0.02  & $<10^{-5}$  & 0.27  & 0.19 & $<10^{-4}$  & $<10^{-4}$\\
			\bottomrule
		\end{tabular}
	}
\end{table}

AFL++ ($\mathtt{cmplog}$ enabled) and SYMCC can also solve these comparison bytes. However, MobFuzz still outperforms AFL++ and SYMCC in 6 out of the 7 projects. The reason is that MobFuzz can optimize other objectives besides the comparison bytes, and AFL++ and SYMCC fail to. MobFuzz can reach a large value of stack memory consumption according to the above experiments. It helps MobFuzz trigger this kind of bug in MAGMA. For example, Bug SQL012 in Table \ref{table_magma_ttb_detail2} is a stack buffer overflow bug. MobFuzz successfully triggers it in 3.2 hours, and AFL++ and SYMCC fail to.


Additionally, the p values in Table \ref{table_magma_bug_p} of MobFuzz are all less than 0.05. However, some results of others are greater than 0.05, e.g., FairFuzz in $\mathtt{libxml2}$. This demonstrates that all the results of MobFuzz are statistically significant.


\subsubsection{Time to Bug}

Table \ref{table_magma_ttb} shows the TTB results, and Table \ref{table_magma_ttb_p} shows the p values of these results. Detailed TTB results of each bug are in Table \ref{table_magma_ttb_detail} and Table \ref{table_magma_ttb_detail2}. MobFuzz outperforms all the baseline fuzzers in 4 out of the 7 projects. The maximum performance improvement is in $\mathtt{poppler}$ compared with AFLFast, which is about 8x. As for the average TTB, MobFuzz has the least TTB compared with others. The reason is that MobFuzz optimizes execution speed and satisfied comparison bytes, and these objectives help to reach less TTB. According to the detailed results, the maximum performance improvement is in Bug PDF007 in Table \ref{table_magma_ttb_detail2}. MobFuzz solves $\mathtt{if( db\to init.busy)}$ and $\mathtt{if( zObj==0 )}$ in 2 minutes. AFLFast succeeds in 22.5 hours.


	\begin{table}[h]
		\setlength{\abovecaptionskip}{0cm}
		\caption{TTB of the fuzzers}
		\label{table_magma_ttb}
		\centering
		\resizebox{0.95\columnwidth}{!}{
			\begin{threeparttable}
			\begin{tabular}{ccccccccc}
				\bottomrule
				\bfseries Targets & \textbf{AFL} & \textbf{AFLFast} & \textbf{AFL++} & \textbf{FairFuzz} & \textbf{honggfuzz} & \textbf{MOPT} & \scriptsize{\textbf{SYMCC}} & \textbf{MobFuzz} \\
				\hline
				libpng & 1.6m\tnote{1}  & 1.7m  & 1.3m  & 1.7m  & 1.5m  & 1.6m & 1.6m  & 1.3m \\
				libtiff & 68.3h\tnote{2}  & 108.0h  & 55.6h  & 105.5h  & 118.1h  & 69.2h & 83.9h  & \textbf{41.1h} \\
				libxml2 & 2.2m  & 2.2m  & 2.2m  & 2.2m  & 53.7m  & 2.0m & 1.8m & \textbf{1.6m} \\
				openssl & 2.5m  & 2.5m  & \textbf{1.6m}  & 2.5m  & 2.4m  & 2.4m & 2.4m  & {1.8m }\\
				php & 4.9m  & 6.2m  & 4.9m  & 6.8m  & 5.7m  & 24.3m & 10.1m & \textbf{4.0m} \\
				poppler & 68.1h  & 79.3h  & 48.3h  & 47.7h  & 64.7h  & 27.9h & 34.7h & \textbf{9.6h} \\
				sqlite3 & 220.9h  & 179.3h  & \textbf{90.7h}  & 164.4h  & 126.0h  & 169.4h & 127.3h & 107.7h \\
				\toprule
				Average & 51.1h  & 52.4h  & 26.9h  & 45.4h  & 44.3h  & 38.1h & 35.2h & \textbf{22.7h} \\
				\bottomrule
			\end{tabular}
			\begin{tablenotes}
				\footnotesize
				\item[1] Smaller values are better. $^2$ ``m'' denotes minutes, and ``h'' denotes hours.
			\end{tablenotes}
		\end{threeparttable}
		}
	\end{table}

Moreover, the p values in Table \ref{table_magma_ttb_p} of MobFuzz are all less than 0.05. However, some results of other fuzzers are greater than 0.05, e.g., AFL in $\mathtt{libtiff}$. This demonstrates that all the TTB results of MobFuzz are statistically significant.

\begin{table}[htb]
	\setlength{\abovecaptionskip}{0cm}
	\caption{p values of the results in Table \ref{table_magma_ttb}}
	\label{table_magma_ttb_p}
	\centering
	\resizebox{0.95\columnwidth}{!}{
		\begin{tabular}{ccccccccc}
			\bottomrule
			\scriptsize{\textbf{Targets}} & \scriptsize{\textbf{AFL}} & \scriptsize{\textbf{AFLFast}} & \scriptsize{\textbf{AFL++}} & \scriptsize{\textbf{FairFuzz}} & \scriptsize{\textbf{honggfuzz}} & \scriptsize{\textbf{MOPT}} & \scriptsize{\textbf{SYMCC}} & \scriptsize{\textbf{MobFuzz}} \\
			\hline 
			libpng & 0.05  & 0.42  & 0.03  & $<10^{-3}$  & $<10^{-4}$  & 0.02 & $<10^{-3}$ & 0.03 \\
			libtiff & 0.12  & $<10^{-3}$  & $<10^{-5}$  & $<10^{-4}$  & 0.01  & $<10^{-3}$ & $<10^{-4}$  & 0.01\\
			libxml2 & $<10^{-5}$  & 0.03  & $<10^{-3}$  & 0.04  & 0.01  & $<10^{-4}$ & 0.29 & 0.03\\
			openssl & 0.33  & $<10^{-4}$  & $<10^{-4}$  & 0.03  & 0.01  & $<10^{-5}$ & $<10^{-4}$ & $<10^{-5}$\\
			php & 0.03  & $<10^{-4}$  & $<10^{-3}$  & $<10^{-5}$  & 0.02 & 0.07 & 0.01 & 0.02\\
			poppler & $<10^{-4}$  & 0.03  & $<10^{-4}$  & 0.07  & $<10^{-3}$  & 0.03 & 0.01 & $<10^{-5}$\\
			sqlite3 & 0.06  & $<10^{-4}$  & 0.02  & $<10^{-4}$  & $<10^{-3}$  & 0.01 & $<10^{-5}$ & 0.04\\
			\bottomrule
		\end{tabular}
	}
\end{table}

In conclusion, in the MAGMA data set, MobFuzz has better bug detection ability and can find the bugs with less TTB compared with the baseline fuzzers.

\section{Discussion}

\subsection{Hyper-parameters}

According to our design in Section \ref{sec_mlmab}, we have 2 parameters in our MPMAB model. In Equation \ref{equ_reward}, $\mathtt{\lambda}$ controls the penalty on objectives that slowdown the fuzzing process. $\mathtt{\gamma}$ in Equation \ref{equ_usb} determines the balance between exploration and exploitation. In this section, we study how the parameters affect the performance of MobFuzz.

\begin{table}[htb]
	\setlength{\abovecaptionskip}{0cm}
	\caption{Average values of objectives of the 12 target programs with different values of $\mathtt{\lambda}$}
	\label{table_lambda}
	\centering
	\resizebox{0.65\columnwidth}{!}{
		\begin{tabular}{cccc}
			\bottomrule
			\bfseries Values of $\mathtt{\lambda}$ & \textbf{Speed} & \textbf{Stack} & \textbf{Cmp} \\
			\toprule
			0.00 & 998.01 & $1.02*10^6$ & $9.96*10^7$ \\
			0.01 & 1019.12 & $9.10*10^5$ & $9.71*10^7$\\
			\textbf{0.10} & 1147.51 & $8.72*10^5$ & $9.60*10^7$\\
			1.00 & 1150.81 & $6.78*10^5$ & $2.38*10^7$\\
			10.00 & 1155.22 & $3.66*10^5$ & $8.51*10^6$\\
			\bottomrule
		\end{tabular}
	}
\end{table}

Table \ref{table_lambda} shows the average values of objectives of the 12 target programs with different values of $\mathtt{\lambda}$. We choose 5 values of $\mathtt{\lambda}$ to show the effect on the performance of MobFuzz. In the $\mathtt{Speed}$ column, the execution speed of fuzzing increases as $\mathtt{\lambda}$ increases. Ideally, we may choose the fastest configuration ($\mathtt{\lambda=10.00}$) because we prefer speed in fuzzing as discussed above. However, when the value of $\mathtt{\lambda}$ is greater than 0.10, the values of stack memory consumption and number of satisfied comparison bytes decrease rapidly, with approximately a {2x} decrease in the $\mathtt{Stack}$ column and a {12x} decrease in the $\mathtt{Cmp}$ column. Therefore, we balance the values of the objectives and choose 0.10 as our configuration of $\mathtt{\lambda}$.

\begin{table}[htb]
	\setlength{\abovecaptionskip}{0cm}
	\caption{Average values of objectives of the 12 target programs with different values of $\mathtt{\gamma}$}
	\label{table_gamma}
	\centering
	\resizebox{0.65\columnwidth}{!}{
		\begin{tabular}{cccc}
			\bottomrule
			\bfseries Values of $\mathtt{\gamma}$ & \textbf{Speed} & \textbf{Stack} & \textbf{Cmp} \\
			\toprule
			0.00 & 512.89 & $1.44*10^6$ & $2.09*10^8$ \\
			\textbf{0.01} & 1147.51 & $8.72*10^5$ & $9.60*10^7$\\
			{0.10} & 1101.97 & $7.21*10^5$ & $7.55*10^7$ \\
			1.00 & 1122.09 & $7.96*10^5$ & $5.63*10^7$\\
			10.00 & 1183.10 & $4.61*10^5$ & $1.07*10^7$\\
			\bottomrule
		\end{tabular}
	}
\end{table}

Table \ref{table_gamma} shows the average values of objectives of the 12 target programs with different values of $\mathtt{\gamma}$. This parameter controls the balance between exploration and exploitation in Equation \ref{equ_usb}. Greater $\mathtt{\gamma}$ means more exploration, and less means more exploitation. We study 5 different values of $\mathtt{\gamma}$ and how $\mathtt{\gamma}$ affects the performance of MobFuzz. When $\mathtt{\gamma}$ is set to 0, the model considers only exploitation, and objectives with greater historical values will be assigned greater scores according to Equation \ref{equ_usb}. In this situation, the number of satisfied comparison bytes will get the greatest score since the $\mathtt{Cmp}$ values are greater than other objective values. Therefore, it can reach the value of $\mathtt{2.09*10^8}$. As $\mathtt{\gamma}$ increases, there will be more exploration. Objectives with smaller values will be assigned greater scores as $\mathtt{\gamma}$ increases. We study different values of $\mathtt{\gamma}$ and the values of the objectives. We finally choose $0.01$ as the configuration. This configuration can produce fast execution speed with appropriate values of stack memory consumption and number of satisfied comparison bytes.


\subsection{Moving to More Objectives}

We choose execution speed, stack memory consumption, and number of satisfied comparison bytes as our objectives in this paper, which is not to say that MobFuzz can handle only three objectives. MobFuzz can be extended to optimize more than 3 objectives with minor changes. For example, if we want to add the number of vulnerable function calls as the 4th objective in MobFuzz, we need to instrument the source code to record the vulnerable functions. Then, we need to modify the number of objectives in the MobFuzz configuration. No further modification is required to optimize these 4 objectives.

\subsection{Threats to Validation}

The randomness in fuzzing is the major threats to validation \cite{bohme2017coverage, ecofuzz, collafl, fairfuzz}. To solve this problem, we conduct repeated experiments to calculate the average values. The p values and $\mathtt{\hat{A}_{12}}$ values are given in our experiments to prove the statistically significant difference. In addition, certain setups of our experiments can be slightly improved. For example, we only set 6 comparison selection strategies in Figure \ref{fig_com_diff}. More strategies could be introduced in the experiments to enrich the comparisons.


\section{Related Work}\label{sec_related}

\subsection{MAB Model in Fuzzing}

The MAB model deals with the problem of optimizing the total reward in finite trials when we are making choices. In fuzzing, there are many situations where we need to maximize the reward. For example, the ultimate goal of fuzzing is to expose as many bugs as possible. Woo et al. \cite{fuzzsim} modeled the parameter configuration as the MAB problem to find more bugs. However, in CGF, the idea of allocating more energy to arms with more bugs will result in triggering the same bugs. Therefore, the number of bugs is not included in our objectives in MobFuzz.

Moreover, Patil et al. \cite{contextual} formalized the process of assigning executions to a test case (energy) as a contextual bandit problem. They proposed a learned model through the policy gradient method to control the energy. Yue et al. \cite{ecofuzz} improved this model and proposed a variant of the adversarial MAB (VAMAB) model. They explained the details of the fuzzing process as a VAMAB model and considered the balance between exploration and exploitation thoroughly. However, in the situation of multiple objectives, we propose our MPMAB model. In contrast to previous work, our MPMAB model deals with the problem of multiple selections combined together. When we have more than one decision to make, e.g., objective combination selection and energy allocation, the classic MAB model is inadequate. Our MLMAB model makes progress in these multiple-selection scenarios.

\subsection{MOO in Fuzzing}
According to our investigation, there are three existing fuzzing tools dealing with multi-objectives in fuzzing. Cerebro uses the idea of the Pareto frontier and non-dominated sorting in \cite{nsga}, as does our NIC algorithm. There are differences between Cerebro and MobFuzz. First, Cerebro does not select objective combinations. As discussed above, there are internal relationships among objectives, which requires us to select the most proper objectives in the current situation. The second difference is whether the optimizing process is an evolutionary procedure. In Cerebro, seeds go through non-dominated sorting, and the Pareto frontier is calculated, which is currently the optimal result. However, this process is only executed once in a fuzzing cycle. We argue that it cannot produce the global optimal solution. Calculating the Pareto frontier and reaching convergence usually require more than 100 iterations through the evolutionary process\cite{nsga}. Additionally, MOOFuzz \cite{moofuzz} has a similar idea to that of Cerebro. Therefore, we argue that it also cannot produce the global optimal solution.

In contrast to MOO in existing fuzzers, MobFuzz selects objective combinations according to the fuzzing state and integrates the evolutionary process into fuzzing without introducing additional overhead according to our evaluation. Based on this, we can produce the global optimal result for multiple objectives without incurring wasted time. In regard to FuzzFactory (two-objective mode) \cite{fuzzfactory}, it is more naive when handling multiple objectives. It uses two continuous $\mathtt{if}$ statements to determine which is better, which can be problematic. For example, it compares seed $\mathtt{X}$ with $\mathtt{Y}$ by $\mathtt{if (A_X>A_Y)\{if(B_X>B_Y)\{prefer\,X\}\}}$ or $\mathtt{if (B_X>B_Y)\{if(A_X>A_Y)\{prefer\,X\}\}}$. Putting objective $\mathtt{A}$ in front of $\mathtt{B}$ or vice versa will lead to incorrect objective optimization. In contrast, MobFuzz produces the optimal value through an evolutionary process and will not fall prey to the above incorrect situations.

\subsection{Power Schedule}
The power schedule of CGF controls the number of mutations and executions on a seed. The original power schedule of AFL allocates more energy than is needed\cite{bohme2017coverage, ecofuzz}. AFLFast \cite{bohme2017coverage} was the pioneering work in improving the power schedule of AFL. It uses a transition probability model to describe the relationship between program paths. AFLFast reduces the energy consumption of AFL through a power schedule and search strategy. Later, Yue et al. proposed EcoFuzz \cite{ecofuzz} to describe the details in the transitions of paths and rewards of seeds through a variant of the adversarial MAB model. Based on the model, the fuzzing process is divided into different states, and different energy values are allocated in these states. Additionally, Entropic \cite{bohme2020boosting} proposes an entropy-based power schedule to allocate more energy to seeds with more information. In contrast, MobFuzz utilizes the MPMAB model to solve two problems, including energy allocation. First, objective combinations are selected. Next, it adaptively allocates appropriate energy to seeds based on the selected combination. Our key contribution is that we extend the power schedule beyond the scope of path coverage. Both AFLFast and EcoFuzz emphasize the path discovery ability of seeds in energy allocation, and information entropy in Entropic is also related to coverage. We design the power schedule to allocate energy based on the objectives we selected. MobFuzz can adjust energy more adaptively based on the current objectives, which broadens the application scenarios of the power schedule in CGF, and this is the key difference compared with previous work.

\subsection{Seed Selection}
In the fuzzing campaign, the fuzzer needs to choose a seed to fuzz when the previous round of fuzzing is finished. It is important to select the best seed in the seed pool based on the goal of the fuzzer. For example, AFL prefers a seed with a faster execution speed and shorter length. When a seed is marked as favored, it will be selected with a higher probability in the next round. Following AFL, MemLock \cite{memlock} and FuzzFactory \cite{fuzzfactory} prefer seeds with more memory consumption and more satisfied comparison bytes, respectively. However, both of them try to optimize only one objective when selecting seeds. As discussed above, there are many situations in which we need to optimize multiple objectives to find deeper bugs. Using only one objective in fuzzing cannot reach the bugs that require multiple triggering conditions. In contrast, we note the lack of consideration and mis-handling of the multiple objectives in CGF. Based on this, we design the MPMAB model to adaptively select the objectives and allocate energy, and we optimize the objectives with NIC. Furthermore, AFLGo \cite{aflgo} and CollAFL \cite{collafl} also select seeds with their specific goals. However, they require complex program analysis to finish the task. Unlike them, MobFuzz does not need additional static analysis to optimize the objectives. IJON \cite{aschermann2020ijon} proposes an annotation mechanism to help the analysts select seeds and guide the fuzzing process. Compared with it, MobFuzz is an automatic fuzzing tool that requires no manual effort to guide the fuzzing process.


\section{Conclusion}
In this paper, we propose MobFuzz to handle the problem of multi-objective optimization in gray-box fuzzing. In MobFuzz, we design a multi-player multi-armed bandit model to adaptively select the objective combinations and allocate energy to seeds. We also propose the NIC algorithm to optimize the objectives without incurring additional performance overhead. Based on the experiments on real-world target programs and the MAGMA data set, we demonstrate the improvements in MobFuzz in contrast to the baseline fuzzers.


\section*{Acknowledgments}

The authors would like to thank the anonymous reviewers for their valuable comments and helpful suggestions. This work is supported by the Natural Science Foundation of China (61902412, 61902405, and 61902416), the Research Project of National University of Defense Technology (ZK20-17 and ZK20-09), the Natural Science Foundation of Hunan Province of China (2021JJ40692), and the National High-level Personnel for Defense Technology Program (2017-JCJQ-ZQ-013).

\bibliographystyle{unsrt}
\bibliography{MobFuzz}

\begin{thebibliography}{10}

\bibitem{bohme2017coverage}
Marcel B{\"o}hme, Van-Thuan Pham, and Abhik Roychoudhury.
\newblock Coverage-based greybox fuzzing as markov chain.
\newblock {\em IEEE Transactions on Software Engineering}, 45(5):489--506,
  2017.

\bibitem{sbst-first}
W.~Miller and D.L. Spooner.
\newblock Automatic generation of floating-point test data.
\newblock {\em IEEE Transactions on Software Engineering}, SE-2(3):223--226,
  1976.

\bibitem{mcminn2011search}
Phil McMinn.
\newblock Search-based software testing: Past, present and future.
\newblock In {\em 2011 IEEE Fourth International Conference on Software
  Testing, Verification and Validation Workshops}, pages 153--163. IEEE, 2011.

\bibitem{sbst-achive}
Mark Harman, Yue Jia, and Yuanyuan Zhang.
\newblock Achievements, open problems and challenges for search based software
  testing.
\newblock In {\em 2015 IEEE 8th International Conference on Software Testing,
  Verification and Validation (ICST)}, pages 1--12, 2015.

\bibitem{sbst-recent}
Xin Yao.
\newblock Some recent work on multi-objective approaches to search-based
  software engineering.
\newblock In G{\"u}nther Ruhe and Yuanyuan Zhang, editors, {\em Search Based
  Software Engineering}, pages 4--15, Berlin, Heidelberg, 2013. Springer Berlin
  Heidelberg.

\bibitem{deb2014multi}
Kalyanmoy Deb.
\newblock Multi-objective optimization.
\newblock In {\em Search methodologies}, pages 403--449. Springer, 2014.

\bibitem{afl}
Michal Zalewski.
\newblock American fuzzy lop.
\newblock In {\em /url{https://lcamtuf.coredump.cx/afl/}}, 2013.

\bibitem{memlock}
Cheng Wen, Haijun Wang, Yuekang Li, Shengchao Qin, Yang Liu, Zhiwu Xu, Hongxu
  Chen, Xiaofei Xie, Geguang Pu, and Ting Liu.
\newblock Memlock: Memory usage guided fuzzing.
\newblock In {\em Proceedings of the ACM/IEEE 42nd International Conference on
  Software Engineering}, ICSE '20, page 765–777, New York, NY, USA, 2020.
  Association for Computing Machinery.

\bibitem{afltech}
Michal Zalewski.
\newblock American fuzzy lop technical details.
\newblock In {\em
  /url{https://lcamtuf.coredump.cx/afl/technical\_datails.txt}}, 2013.

\bibitem{ecofuzz}
Tai Yue, Pengfei Wang, Yong Tang, Enze Wang, Bo~Yu, Kai Lu, and Xu~Zhou.
\newblock Ecofuzz: Adaptive energy-saving greybox fuzzing as a variant of the
  adversarial multi-armed bandit.
\newblock In {\em 29th {USENIX} Security Symposium ({USENIX} Security 20)},
  pages 2307--2324. {USENIX} Association, August 2020.

\bibitem{cerebro}
Yuekang Li, Yinxing Xue, Hongxu Chen, Xiuheng Wu, Cen Zhang, Xiaofei Xie,
  Haijun Wang, and Yang Liu.
\newblock Cerebro: Context-aware adaptive fuzzing for effective vulnerability
  detection.
\newblock In {\em Proceedings of the 2019 27th ACM Joint Meeting on European
  Software Engineering Conference and Symposium on the Foundations of Software
  Engineering}, ESEC/FSE 2019, page 533–544, New York, NY, USA, 2019.
  Association for Computing Machinery.

\bibitem{nsga}
K.~{Deb}, A.~{Pratap}, S.~{Agarwal}, and T.~{Meyarivan}.
\newblock A fast and elitist multiobjective genetic algorithm: Nsga-ii.
\newblock {\em IEEE Transactions on Evolutionary Computation}, 6(2):182--197,
  2002.

\bibitem{mab}
P.~Whittle.
\newblock Multi-armed bandits and the gittins index.
\newblock {\em Journal of the Royal Statistical Society. Series B
  (Methodological)}, 42(2):143--149, 1980.

\bibitem{2021Reinforcement}
J.~Wang, C.~Song, and H.~Yin.
\newblock Reinforcement learning-based hierarchical seed scheduling for greybox
  fuzzing.
\newblock In {\em Network and Distributed System Security Symposium}, 2021.

\bibitem{li2020unifuzz}
Yuwei Li, Shouling Ji, Yuan Chen, Sizhuang Liang, Wei-Han Lee, Yueyao Chen,
  Chenyang Lyu, Chunming Wu, Raheem Beyah, Peng Cheng, Kangjie Lu, and Ting
  Wang.
\newblock Unifuzz: A holistic and pragmatic metrics-driven platform for
  evaluating fuzzers, 2020.

\bibitem{tortoisefuzz}
Yanhao Wang, Xiangkun Jia, Yuwei Liu, Kyle Zeng, Tiffany Bao, Dinghao Wu, and
  Purui Su.
\newblock Not all coverage measurements are equal: Fuzzing by coverage
  accounting for input prioritization.
\newblock NDSS, 2020.

\bibitem{fairfuzz}
Caroline Lemieux and Koushik Sen.
\newblock Fairfuzz: A targeted mutation strategy for increasing greybox fuzz
  testing coverage.
\newblock In {\em Proceedings of the 33rd ACM/IEEE International Conference on
  Automated Software Engineering}, pages 475--485, 2018.

\bibitem{ptfuzz}
G.~{Zhang}, X.~{Zhou}, Y.~{Luo}, X.~{Wu}, and E.~{Min}.
\newblock Ptfuzz: Guided fuzzing with processor trace feedback.
\newblock {\em IEEE Access}, 6:37302--37313, 2018.

\bibitem{aflbug}
Michal Zalewski.
\newblock Afl vulnerability trophy case.
\newblock In {\em /url{https://lcamtuf.coredump.cx/afl/bugs}}, 2013.

\bibitem{fuzzfactory}
Rohan Padhye, Caroline Lemieux, Koushik Sen, Laurent Simon, and Hayawardh
  Vijayakumar.
\newblock Fuzzfactory: Domain-specific fuzzing with waypoints.
\newblock {\em Proc. ACM Program. Lang.}, 3(OOPSLA), October 2019.

\bibitem{bubeck2012regret}
Sébastien Bubeck and Nicolò Cesa-Bianchi.
\newblock Regret analysis of stochastic and nonstochastic multi-armed bandit
  problems, 2012.

\bibitem{patil2018greybox}
Ketan Patil and Aditya Kanade.
\newblock Greybox fuzzing as a contextual bandits problem, 2018.

\bibitem{aflhier}
Jinghan Wang, Chengyu Song, and Heng Yin.
\newblock Reinforcement learning-based hierarchical seed scheduling for greybox
  fuzzing.
\newblock 01 2021.

\bibitem{zeror}
Chijin Zhou, Mingzhe Wang, Jie Liang, Zhe Liu, and Yu~Jiang.
\newblock Zeror: Speed up fuzzing with coverage-sensitive tracing and
  scheduling.
\newblock In {\em 2020 35th IEEE/ACM International Conference on Automated
  Software Engineering (ASE)}, pages 858--870, 2020.

\bibitem{fullspeed}
Stefan Nagy and Matthew Hicks.
\newblock Full-speed fuzzing: Reducing fuzzing overhead through coverage-guided
  tracing.
\newblock In {\em 2019 IEEE Symposium on Security and Privacy (SP)}, pages
  787--802. IEEE, 2019.

\bibitem{instrcr}
Cao Zhang, Wei~Yu Dong, and Yu~Zhu Ren.
\newblock Instrcr: Lightweight instrumentation optimization based on
  coverage-guided fuzz testing.
\newblock In {\em 2019 IEEE 2nd International Conference on Computer and
  Communication Engineering Technology (CCET)}, pages 74--78. IEEE, 2019.

\bibitem{agrawal1995sample}
Rajeev Agrawal.
\newblock Sample mean based index policies with o (log n) regret for the
  multi-armed bandit problem.
\newblock {\em Advances in Applied Probability}, pages 1054--1078, 1995.

\bibitem{bohme2020boosting}
Marcel B{\"o}hme, Valentin~JM Man{\`e}s, and Sang~Kil Cha.
\newblock Boosting fuzzer efficiency: An information theoretic perspective.
\newblock In {\em Proceedings of the 28th ACM Joint Meeting on European
  Software Engineering Conference and Symposium on the Foundations of Software
  Engineering}, pages 678--689, 2020.

\bibitem{klees2018evaluating}
George Klees, Andrew Ruef, Benji Cooper, Shiyi Wei, and Michael Hicks.
\newblock Evaluating fuzz testing.
\newblock In {\em Proceedings of the 2018 ACM SIGSAC Conference on Computer and
  Communications Security}, pages 2123--2138, 2018.

\bibitem{hazimeh2020magma}
Ahmad Hazimeh, Adrian Herrera, and Mathias Payer.
\newblock Magma: A ground-truth fuzzing benchmark.
\newblock {\em Proceedings of the ACM on Measurement and Analysis of Computing
  Systems}, 4(3):1--29, 2020.

\bibitem{afl++}
Andrea Fioraldi, Dominik Maier, Heiko Ei{\ss}feldt, and Marc Heuse.
\newblock Afl++: Combining incremental steps of fuzzing research.
\newblock In {\em 14th $\{$USENIX$\}$ Workshop on Offensive Technologies
  ($\{$WOOT$\}$ 20)}, 2020.

\bibitem{honggfuzz}
Google.
\newblock honggfuzz: Security oriented software fuzzer. supports evolutionary,
  feedback-driven fuzzing based on code coverage (sw and hw based).
\newblock In {\em /url{https://honggfuzz.dev/}}, 2015.

\bibitem{mopt}
Chenyang Lyu, Shouling Ji, Chao Zhang, Yuwei Li, Wei-Han Lee, Yu~Song, and
  Raheem Beyah.
\newblock {MOPT}: Optimized mutation scheduling for fuzzers.
\newblock In {\em 28th {USENIX} Security Symposium ({USENIX} Security 19)},
  pages 1949--1966, Santa Clara, CA, August 2019. {USENIX} Association.

\bibitem{poeplau2020symbolic}
Sebastian Poeplau and Aur{\'e}lien Francillon.
\newblock Symbolic execution with symcc: Don't interpret, compile!
\newblock In {\em 29th $\{$USENIX$\}$ Security Symposium ($\{$USENIX$\}$
  Security 20)}, pages 181--198, 2020.

\bibitem{collafl}
S.~{Gan}, C.~{Zhang}, X.~{Qin}, X.~{Tu}, K.~{Li}, Z.~{Pei}, and Z.~{Chen}.
\newblock Collafl: Path sensitive fuzzing.
\newblock In {\em 2018 IEEE Symposium on Security and Privacy (SP)}, pages
  679--696, 2018.

\bibitem{fuzzsim}
Maverick Woo, Sang~Kil Cha, Samantha Gottlieb, and David Brumley.
\newblock Scheduling black-box mutational fuzzing.
\newblock In {\em Proceedings of the 2013 ACM SIGSAC Conference on Computer
  Communications Security}, CCS '13, page 511–522, New York, NY, USA, 2013.
  Association for Computing Machinery.

\bibitem{contextual}
Ketan Patil and Aditya Kanade.
\newblock Greybox fuzzing as a contextual bandits problem.
\newblock {\em CoRR}, abs/1806.03806, 2018.

\bibitem{moofuzz}
Xiaoqi Zhao, Haipeng Qu, Wenjie Lv, Shuo Li, and Jianliang Xu.
\newblock Moofuzz: Many-objective optimization seed schedule for fuzzer.
\newblock {\em Mathematics}, 9(3), 2021.

\bibitem{aflgo}
Marcel B\"{o}hme, Van-Thuan Pham, Manh-Dung Nguyen, and Abhik Roychoudhury.
\newblock Directed greybox fuzzing.
\newblock In {\em Proceedings of the 2017 ACM SIGSAC Conference on Computer and
  Communications Security}, CCS '17, page 2329–2344, New York, NY, USA, 2017.
  Association for Computing Machinery.

\bibitem{aschermann2020ijon}
Cornelius Aschermann, Sergej Schumilo, Ali Abbasi, and Thorsten Holz.
\newblock Ijon: Exploring deep state spaces via fuzzing.
\newblock In {\em 2020 IEEE Symposium on Security and Privacy (SP)}, pages
  1597--1612. IEEE, 2020.

\end{thebibliography}

\appendix

\section{Appendix}

\subsection{Evaluation of the MPMAB Model}

\begin{table}[h]
	\caption{12 real-world programs and related state-of-the-art papers}
	\label{table_target_use}
	\centering
	\resizebox{\columnwidth}{!}{
		\begin{tabular}{ll}
			\bottomrule
			\bfseries Programs & \textbf{Papers and publication} \\
			\toprule
			avconv & AFLSMART (TSE2019), MOPT (Security2019), Intriguer (CCS2019) \\
			exiv2 & CollAFL (S\&P2018), SLF (ICSE2019), UNIFUZZ (Security2021)\\
			infotocap & MOPT (Security2019), EcoFuzz (Security2020), UNIFUZZ (Security2021) \\
			mp42aac & MOPT (Security2019), MemLock (ICSE2020), UNIFUZZ (Security2021)\\
			mp4tag & MemLock (ICSE2020), PANGOLIN (S\&P2020), Fuzzguard (Security2020 \\
			nm & Angora (S\&P2018), FuZZan (Security2020), EcoFuzz (Security2020) \\
			pdfinfo & MOPT (Security2019), ProFuzzer (S\&P2019), Fuzzguard (Security2020)  \\
			readelf & Fuzzification (Security2019), MEUZZ (RAID2020), OptiMin (ISSTA2021) \\
			tiff2pdf & Steelix (FSE2017), GREYONE (Security2020), Intriguer (CCS2019) \\
			tiff2ps & QSYM (Security2018), Matryoshka (CCS2019), MEUZZ (RAID2020) \\
			txtext & ProFuzzer (S\&P2019), Fuzzguard (Security2020), Ferry (Security2022) \\
			xmllint & AFLFast (CCS2016), Matryoshka (CCS2019), EcoFuzz (Security2020)\\
			\bottomrule
		\end{tabular}
	}
\end{table}

Table \ref{table_target_use} lists the 12 real-world programs used in our experiments and the related papers which used the programs.

	\begin{figure}
		\centering
		\includegraphics[width=0.8\columnwidth]{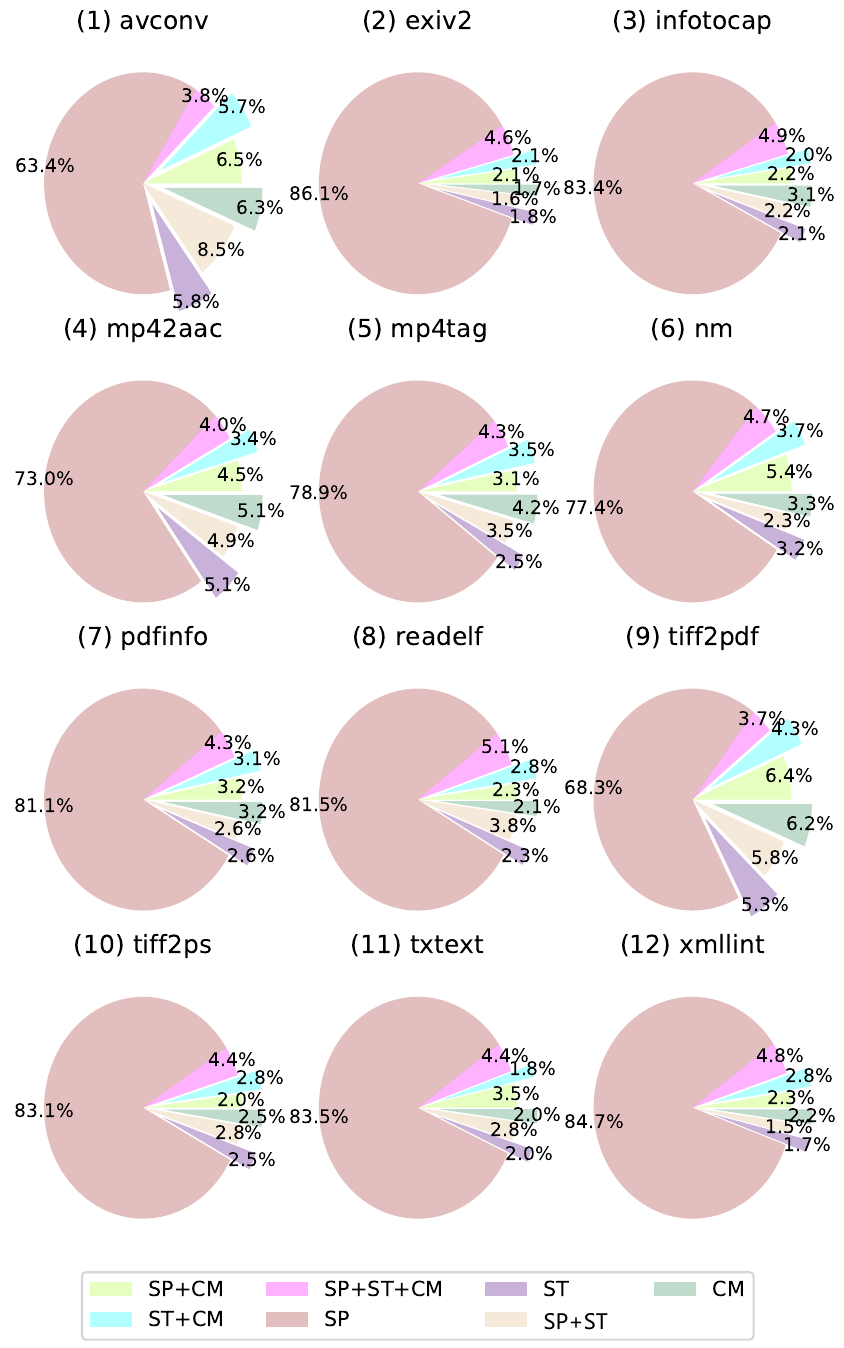}
		\caption{\footnotesize{Percentages of the chosen objective combinations. Different colors indicate different objective combinations.}}
		\label{fig_com_sel}
	\end{figure}

\textbf{Objective combination selection. }Figure \ref{fig_com_sel} shows the percentages of the chosen objective combinations. Different colors indicate different objective combinations.

	\begin{figure}
		\centering
		\includegraphics[width=0.8\columnwidth]{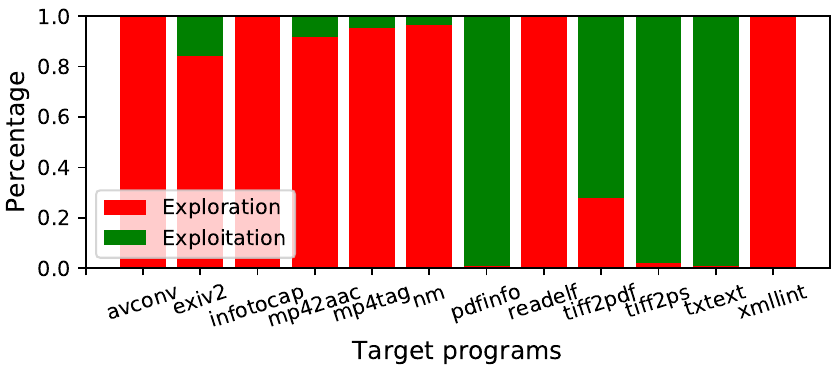}
		\caption{\footnotesize{Percentages of exploration and exploitation states during fuzzing.}}
		\label{fig_fuzzing_state}
	\end{figure}

\textbf{Power Schedule. } Figure \ref{fig_fuzzing_state} demonstrates the percentages of exploration and exploitation states during fuzzing. Exploration and exploitation can transfer to each other depending on whether new seeds are generated. As we can see from the figure, the percentages of different target programs are completely different. The number of total paths of the programs differs. Therefore, the transformations between exploration and exploitation are different. MobFuzz discovers new paths more easily in some programs, and the state transfers to the exploration state. If no new seeds are generated, the exploitation state begins.

\subsection{Evaluation on NIC}

	\begin{figure}
		\centering
		\includegraphics[width=0.8\columnwidth]{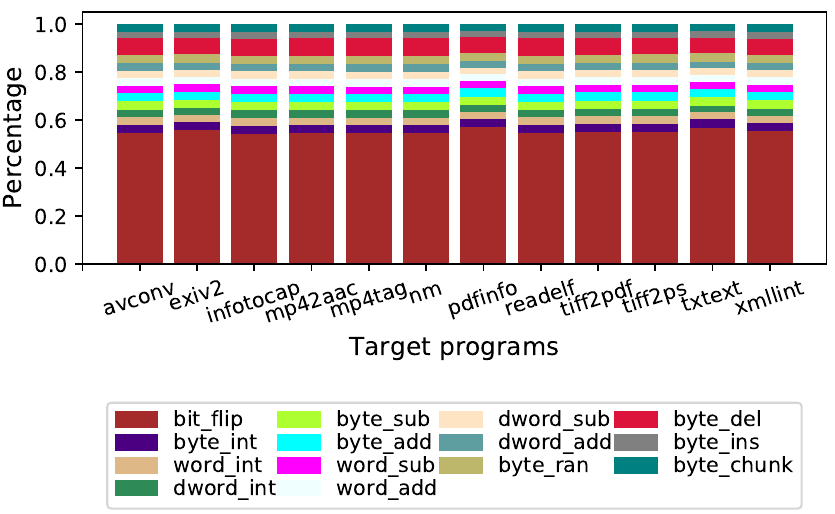}
		\caption{\footnotesize{Percentages of mutation operators during fuzzing.}}
		\label{fig_muter_score}
	\end{figure}

We propose our new strategy by selecting the mutation operators with better historical performance. Figure \ref{fig_muter_score} shows the proportion of the selected operators. We can see that flipping 1 bit of the seed covers more than 50\% of the figure, which demonstrates the simplicity and effectiveness of the $\mathtt{bitflip}$ operator.

	\begin{figure}
		\centering
		\includegraphics[width=0.8\columnwidth]{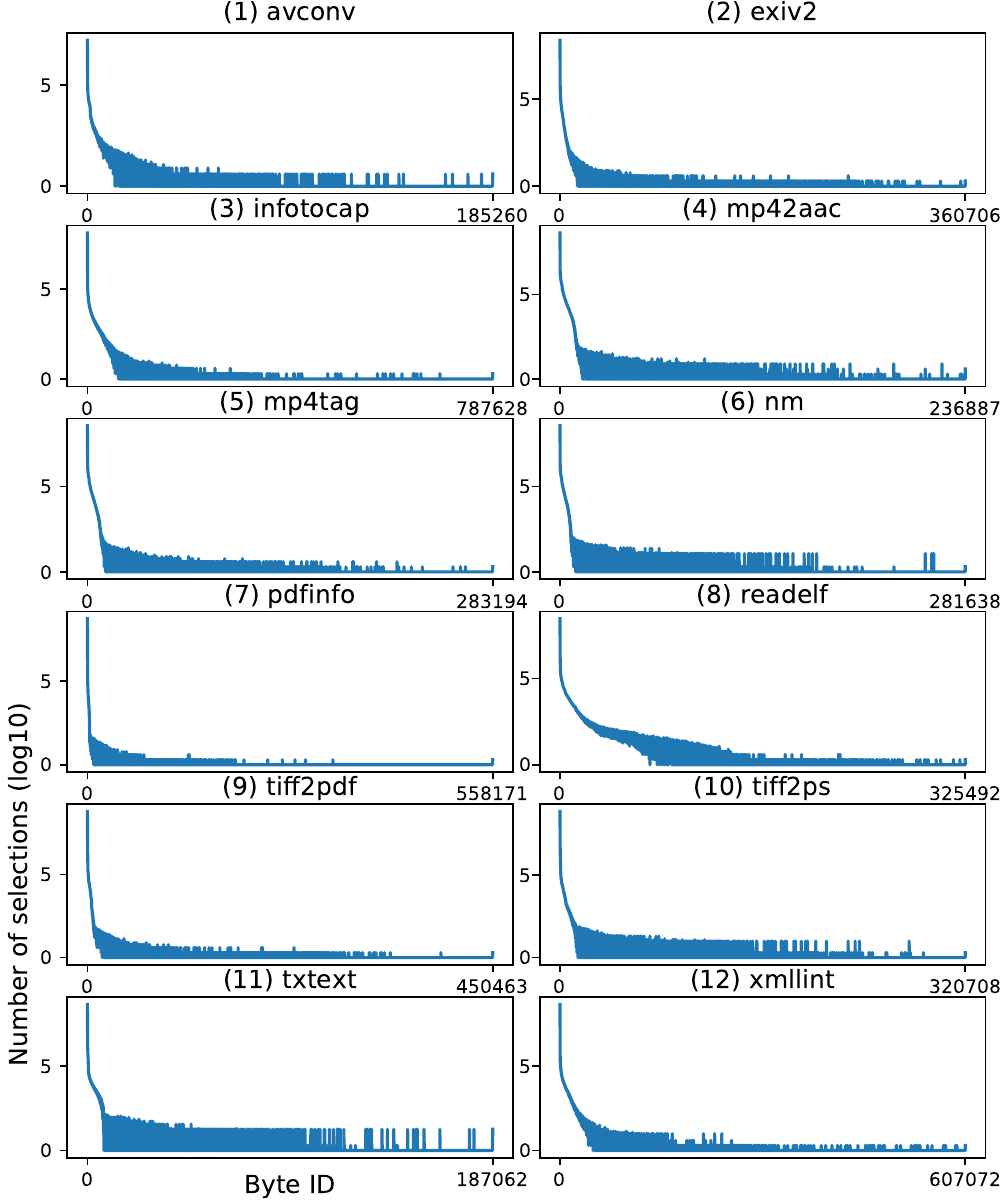}
		\caption{\footnotesize{The number of times a byte is selected.}}
		\label{fig_gene_score}
	\end{figure}

Figure \ref{fig_gene_score} shows the number of selections on the bytes of seeds. (10, 999) indicates that the 10th byte of the input is selected 999 times by the mutation operator. We also tend to select the bytes with better historical performance in our design. In the figure, bytes in the front positions of the seed are more likely to be chosen, which implies that mutating these bytes can bring more reward to the fuzzing process. For example, the header of a JPEG file contains structural information about the file. The header bytes are more important than the subsequent bytes.

\subsection{Bugs in the 12 Real-world Programs}

	\begin{table}[h]
		\caption{Bugs discovered by MobFuzz in the 12 real-world programs}
		\label{table_new_bug}
		\centering
		\resizebox{\columnwidth}{!}{
			\begin{tabular}{ll}
				\bottomrule
				\bfseries Programs & \textbf{Bug description} \\
				\toprule
				infotocap & heap-buffer-overflow in captoinfo.c in \_nc\_infotocap() \\
				infotocap & global-buffer-overflow in comp\_hash.c in \_nc\_find\_entry()\\
				infotocap & stack-buffer-overflow in infotocap \\
				mp42aac & heap-buffer-overflow in Ap4TrunAtom.cpp in AP4\_TrunAtom()\\
				mp42aac & heap-buffer-overflow in Ap4Utils.cpp in SkipBits() \\
				mp42aac & heap-buffer-overflow in Ap4Dec3Atom.cpp in AP4\_Dec3Atom() \\
				mp4tag & heap-buffer-overflow in Ap4ByteStream.cpp in WritePartial() \\
				mp4tag & heap-buffer-overflow in Ap4RtpAtom.cpp in AP4\_RtpAtom() \\
				mp4tag & heap-buffer-overflow in Ap4Utils.h in AP4\_BytesToUInt32BE \\
				tiff2pdf & heap-buffer-overflow in t2p\_read\_tiff\_size() \\
				tiff2ps & heap-buffer-overflow in tiff2ps.c in PSDataColorContig() \\
				\bottomrule
			\end{tabular}
		}
	\end{table}

Table \ref{table_new_bug} shows the real bugs discovered by MobFuzz in the 12 real-world programs. These bugs are collected from the unique crashes with the help of ASAN and GDB. In this table, most of the bugs are overflow bugs. From this result, we can see the vulnerability detection ability of MobFuzz. Moreover, MobFuzz can detect more unique bugs than the baseline fuzzers.

\subsection{Detailed TTB in the MAGMA Data Set}

Table \ref{table_magma_ttb_detail} and Table \ref{table_magma_ttb_detail2} show the detailed TTB of each bug in the MAGMA data set. In total, there are 118 bugs in the data set. The identification of each bug is in the format of ``Project+Number", e.g., ``PNG001" denotes the 1st bug in the libpng project. In addition, ``PNG" denotes libpng, ``TIF" denotes libtiff, ``XML" denotes libxml2, ``SSL" denotes openssl, ``PHP" denotes php, ``PDF" denotes poppler, and ``SQL" denotes splite3.

\begin{table*}
	\setlength{\abovecaptionskip}{0cm}
	\caption{Detailed TTB in the MAGMA data set}
	\label{table_magma_ttb_detail}
	\centering
	\begin{threeparttable}
		\begin{tabular}{ccccccccc}
			\bottomrule
			\bfseries Bug ID & \textbf{AFL} & \textbf{AFLFast} & \textbf{AFL++} & \textbf{FairFuzz} & \textbf{honggfuzz} & \textbf{MOPT} &\textbf{SYMCC} & \textbf{MobFuzz} \\
			\hline 
			PNG001 & 15.0s\tnote{1} & 15.0s & 10.0s & 15.0s & 15.0s & 15.0s & 15.0s & 15.0s\\ 
			PNG002 & $\times$\tnote{2} & $\times$ & $\times$ & $\times$ & $\times$ & $\times$ & $\times$ & $\times$\\ 
			PNG003 & 20.0s & 20.0s & 15.0s & 20.0s & 15.0s & 15.0s & 20.0s & 10.0s\\ 
			PNG004 & 15.0s & 20.0s & 15.0s & 20.0s & 20.0s & 20.0s & 20.0s & 15.0s\\ 
			PNG005 & 15.0s & 15.0s & 15.0s & 15.0s & 15.0s & 15.0s & 15.0s & 10.0s\\ 
			PNG006 & 15.0s & 15.0s & 10.0s & 15.0s & 10.0s & 15.0s & 10.0s & 15.0s\\ 
			PNG007 & 15.0s & 15.0s & 15.0s & 15.0s & 15.0s & 15.0s & 15.0s & 10.0s\\ 
			\toprule
			TIF001 & 23.3h & 21.9h & $\times$ & $\times$ & $\times$ & $\times$ & $\times$ & 4.2h\\ 
			TIF002 & 7.0m & $\times$ & 5.0h & 8.3h & $\times$ & 4.5h & 13.6h & 2.2h\\ 
			TIF003 & 2.0m & 15.0s & 15.0s & 15.0s & 15.0s & 45.0s & 15.0s & 10.0s\\ 
			TIF004 & $\times$ & $\times$ & $\times$ & $\times$ & $\times$ & $\times$ & $\times$ & $\times$\\ 
			TIF005 & $\times$ & $\times$ & 180 & $\times$ & 180 & $\times$ & $\times$ & $\times$\\ 
			TIF006 & 9.0h & $\times$ & 4.0m & $\times$ & $\times$ & 11.8h & 12.7h & 3.0m\\ 
			TIF007 & 1.0m & 50.0s & 15.0s & 1.0m & 15.0s & 15.0s & 15.0s & 15.0s\\ 
			TIF008 & 3.8h & 6.0h & $\times$ & $\times$ & $\times$ & 4.0h & 7.1h & 3.1h\\ 
			TIF009 & 6.7h & 7.0h & 2.4h & 1.0h & 22.5h & 2.0h & 2.1h & 8.3h\\ 
			TIF010 & 1.1h & 1.0h & 29.0m & 27.0m & 5.0m & 13.0m & 29.2m & 4.0m\\ 
			TIF011 & $\times$ & $\times$ & $\times$ & $\times$ & $\times$ & $\times$ & $\times$ & $\times$\\ 
			TIF012 & 10.0s & 10.0s & 10.0s & 10.0s & 10.0s & 10.0s & 10.0s & 5.0s\\ 
			TIF013 & $\times$ & $\times$ & $\times$ & $\times$ & $\times$ & $\times$ & $\times$ & $\times$\\ 
			TIF014 & 1.0m & 50.0s & 15.0s & 1.0m & 15.0s & 15.0s & 15.0s & 15.0s\\ 
			\toprule
			XML001 & 15.0s & 15.0s & 20.0s & 15.0s & 10.0s & 15.0s & 10.0s & 15.0s\\ 
			XML002 & $\times$ & $\times$ & $\times$ & $\times$ & $\times$ & $\times$ & $\times$ & $\times$\\ 
			XML003 & 15.0s & 15.0s & 10.0s & 15.0s & 15.0s & 15.0s & 15.0s & 15.0s\\ 
			XML004 & $\times$ & $\times$ & $\times$ & $\times$ & $\times$ & $\times$ & $\times$ & $\times$\\ 
			XML005 & $\times$ & $\times$ & $\times$ & $\times$ & $\times$ & $\times$ & $\times$ & $\times$\\ 
			XML006 & 15.0s & 15.0s & 20.0s & 15.0s & 20.0s & 15.0s & 15.0s & 10.0s\\ 
			XML007 & $\times$ & $\times$ & $\times$ & $\times$ & $\times$ & $\times$ & $\times$ & $\times$\\ 
			XML008 & 22.0s & 20.0s & 15.0s & 20.0s & 15.0s & 10.0s & 10.0s & 15.0s\\ 
			XML009 & 15.0s & 15.0s & 10.0s & 15.0s & 15.0s & 15.0s & 15.0s & 5.0s\\ 
			XML010 & $\times$ & $\times$ & $\times$ & $\times$ & $\times$ & $\times$ & $\times$ & $\times$\\ 
			XML011 & 15.0s & 15.0s & 20.0s & 15.0s & 52.0m & 15.0s & 15.0s & 10.0s\\ 
			XML012 & 15.0s & 15.0s & 20.0s & 15.0s & 10.0s & 15.0s & 15.0s & 15.0s\\ 
			XML013 & $\times$ & $\times$ & $\times$ & $\times$ & $\times$ & $\times$ & $\times$ & $\times$\\ 
			XML014 & $\times$ & $\times$ & $\times$ & $\times$ & $\times$ & $\times$ & $\times$ & $\times$\\ 
			XML015 & $\times$ & $\times$ & $\times$ & $\times$ & $\times$ & $\times$ & $\times$ & $\times$\\ 
			XML016 & $\times$ & $\times$ & $\times$ & $\times$ & $\times$ & $\times$ & $\times$ & $\times$\\ 
			XML017 & 20.0s & 20.0s & 15.0s & 20.0s & 15.0s & 20.0s & 15.0s & 10.0s\\ 
			XML018 & $\times$ & $\times$ & $\times$ & $\times$ & $\times$ & $\times$ & $\times$ & $\times$\\ 
			\toprule
			SSL001 & 20.0s & 20.0s & 15.0s & 20.0s & 20.0s & 20.0s & 20.0s & 20.0s\\ 
			SSL002 & 15.0s & 15.0s & 15.0s & 15.0s & 15.0s & 15.0s & 15.0s & 15.0s\\ 
			SSL003 & 20.0s & 20.0s & 20.0s & 20.0s & 15.0s & 15.0s & 15.0s & 10.0s\\ 
			SSL004 & $\times$ & $\times$ & $\times$ & $\times$ & $\times$ & $\times$ & $\times$ & $\times$\\ 
			SSL005 & 15.0s & 15.0s & 15.0s & 15.0s & 15.0s & 15.0s & 15.0s & 5.0s\\ 
			SSL006 & $\times$ & $\times$ & $\times$ & $\times$ & $\times$ & $\times$ & $\times$ & $\times$\\ 
			SSL007 & $\times$ & $\times$ & $\times$ & $\times$ & $\times$ & $\times$ & $\times$ & $\times$\\ 
			SSL008 & 15.0s & 15.0s & 20.0s & 15.0s & 15.0s & 15.0s & 15.0s & 10.0s\\ 
			SSL009 & 15.0s & 15.0s & 15.0s & 15.0s & 10.0s & 15.0s & 15.0s & 15.0s\\ 
			SSL010 & 10.0s & 10.0s & 10.0s & 10.0s & 10.0s & 10.0s & 10.0s & 5.0s\\ 
			SSL011 & $\times$ & $\times$ & $\times$ & $\times$ & $\times$ & $\times$ & $\times$ & $\times$\\ 
			SSL012 & $\times$ & $\times$ & $\times$ & $\times$ & $\times$ & $\times$ & $\times$ & $\times$\\ 
			SSL013 & $\times$ & $\times$ & $\times$ & $\times$ & $\times$ & $\times$ & $\times$ & $\times$\\ 
			SSL014 & $\times$ & $\times$ & $\times$ & $\times$ & $\times$ & $\times$ & $\times$ & $\times$\\ 
			SSL015 & $\times$ & $\times$ & $\times$ & $\times$ & $\times$ & $\times$ & $\times$ & $\times$\\ 
			SSL016 & 15.0s & 15.0s & 15.0s & 15.0s & 15.0s & 15.0s & 15.0s & 10.0s\\ 
			SSL017 & $\times$ & $\times$ & $\times$ & $\times$ & $\times$ & $\times$ & $\times$ & $\times$\\ 
			SSL018 & $\times$ & $\times$ & $\times$ & $\times$ & $\times$ & $\times$ & $\times$ & $\times$\\ 
			SSL019 & 10.0s & 10.0s & 10.0s & 10.0s & 10.0s & 10.0s & 10.0s & 10.0s\\ 
			SSL020 & 15.0s & 15.0s & 15.0s & 15.0s & 20.0s & 15.0s & 15.0s & 10.0s\\ 
			SSL021 & $\times$ & $\times$ & $\times$ & $\times$ & $\times$ & $\times$ & $\times$ & $\times$\\ 
			SSL022 & $\times$ & $\times$ & $\times$ & $\times$ & $\times$ & $\times$ & $\times$ & $\times$\\ 
			\bottomrule
		\end{tabular}
		\begin{tablenotes}
			\footnotesize
			\item[1] ``s'' denotes seconds, ``m'' denotes minutes, and ``h'' denotes hours. $^2$ ``$\times$" denotes the fuzzer cannot find the bug.
		\end{tablenotes}
	\end{threeparttable}
\end{table*}

\begin{table*}
	\setlength{\abovecaptionskip}{0cm}
	\caption{Detailed TTB in the MAGMA data set}
	\label{table_magma_ttb_detail2}
	\centering
		\begin{tabular}{ccccccccc}
			\bottomrule
			\bfseries Bug ID & \textbf{AFL} & \textbf{AFLFast} & \textbf{AFL++} & \textbf{FairFuzz} & \textbf{honggfuzz} & \textbf{MOPT} & \textbf{SYMCC} & \textbf{MobFuzz} \\
			\hline
			PHP001 & $\times$ & $\times$ & $\times$ & $\times$ & $\times$ & $\times$ & $\times$ & $\times$\\ 
			PHP002 & 15.0s & 15.0s & 10.0s & 15.0s & 10.0s & 15.0s & 15.0s & 10.0s\\ 
			PHP003 & 15.0s & 15.0s & 15.0s & 15.0s & 15.0s & 15.0s & 15.0s &  15.0s\\ 
			PHP004 & 15.0s & 15.0s & 15.0s & 15.0s & 10.0s & 15.0s & 15.0s & 5.0s\\ 
			PHP005 & $\times$ & $\times$ & $\times$ & $\times$ & $\times$ & $\times$ & $\times$ & $\times$\\ 
			PHP006 & 15.0s & 17.0s & 15.0s & 10.0s & 10.0s & 20.0s & 10.0s & 10.0s\\ 
			PHP007 & $\times$ & $\times$ & $\times$ & $\times$ & $\times$ & $\times$ & $\times$ & $\times$\\ 
			PHP008 & $\times$ & $\times$ & $\times$ & $\times$ & $\times$ & $\times$ & $\times$ & $\times$\\ 
			PHP009 & 15.0s & 15.0s & 15.0s & 15.0s & 15.0s & 15.0s & 15.0s & 10.0s\\ 
			PHP010 & $\times$ & $\times$ & $\times$ & $\times$ & $\times$ & $\times$ & $\times$ & $\times$\\ 
			PHP011 & 10.0s & 10.0s & 15.0s & 10.0s & 10.0s & 10.0s & 10.0s & 10.0s\\ 
			PHP012 & $\times$ & $\times$ & $\times$ & $\times$ & $\times$ & $\times$ & $\times$ & $\times$\\ 
			PHP013 & $\times$ & $\times$ & $\times$ & $\times$ & $\times$ & $\times$ & $\times$ & $\times$\\ 
			PHP014 & $\times$ & $\times$ & $\times$ & $\times$ & $\times$ & $\times$ & $\times$ & $\times$\\ 
			PHP015 & 3.0m & 4.0m & 3.0m & 5.0m & 4.0m & 22.0m & 8.0m & 2.5m\\ 
			PHP016 & $\times$ & $\times$ & $\times$ & $\times$ & $\times$ & $\times$ & $\times$ & $\times$\\ 
			PHP017 & 10.0s & 15.0s & 10.0s & 10.0s & 10.0s & 15.0s & 15.0s & 15.0s\\ 
			PHP018 & $\times$ & $\times$ & $\times$ & $\times$ & $\times$ & $\times$ & $\times$ & $\times$\\ 
			PHP019 & $\times$ & $\times$ & $\times$ & $\times$ & $\times$ & $\times$ & $\times$ & $\times$\\ 
			PHP020 & 10.0s & 15.0s & 10.0s & 10.0s & 10.0s & 15.0s & 15.0s & 10.0s\\ 
			PHP021 & 10.0s & 15.0s & 10.0s & 10.0s & 10.0s & 15.0s & 15.0s & 5.0s\\ 
			\toprule
			PDF001 & 40.0s & 35.0s & 20.0s & 35.0s & 55.0s & 35.0s & 25.0s & 15.0s\\ 
			PDF002 & 3.3h & 22.0m & 48.0m & 2.9h & 15.2h & 2.0m & 1.9h & 5.5h\\ 
			PDF003 & 20.0s & 15.0s & 20.0s & 15.0s & 25.0s & 15.0s & 15.0s & 10.0s\\ 
			PDF004 & $\times$ & $\times$ & $\times$ & $\times$ & $\times$ & $\times$ & $\times$ & $\times$\\ 
			PDF005 & 25.0s & 20.0s & 25.0s & 20.0s & 60.0s & 20.0s & 20.0s & 15.0s\\ 
			PDF006 & 30.0s & 25.0s & 30.0s & 25.0s & 20.0s & 25.0s & 20.0s & 25.0s\\ 
			PDF007 & 2.0h & 22.5h & 45.0m & 3.0h & 2.5m & 4.0m & 1.3h & 2.0m\\ 
			PDF008 & 25.0s & 25.0s & 15.0s & 25.0s & 25.0s & 25.0s & 20.0s & 25.0s\\ 
			PDF009 & 5.1h & 55.0m & 41.0m & 40.0m & 32.0m & 50.0m & 1.1h & 30.0m\\ 
			PDF010 & 7.1h & 4.8h & 1.0h & 3.2h & $\times$ & 41.0m & 1.3h & 29.0m\\ 
			PDF011 & 15.0s & 15.0s & 15.0s & 15.0s & 15.0s & 15.0s & 15.0s & 10.0s\\ 
			PDF012 & 15.0s & 15.0s & 15.0s & 15.0s & 15.0s & 15.0s & 15.0s & 15.0s\\ 
			PDF013 & $\times$ & $\times$ & $\times$ & $\times$ & $\times$ & $\times$ & $\times$ & $\times$\\ 
			PDF014 & 20.8h & 21.1h & 3.2h & 7.0h & 51.0h & 14.7h & 3.3h & 28.0m\\ 
			PDF015 & $\times$ & $\times$ & $\times$ & $\times$ & $\times$ & $\times$ & $\times$ & $\times$\\ 
			PDF016 & 3.6h & 1.0h & 1.0h & 3.4h & 3.3m & 4.0m & 41.7m & 2.0h\\ 
			PDF017 & $\times$ & $\times$ & $\times$ & $\times$ & $\times$ & $\times$ & $\times$ & $\times$\\ 
			PDF018 & $\times$ & $\times$ & 11.5h & $\times$ & $\times$ & 12.0h & $\times$ & 1.0h\\ 
			PDF019 & 4.4h & 6.3h & 2.0h & 4.0h & 12.0m & 8.0m & 1.4h & 7.0m\\ 
			PDF020 & 15.0s & 15.0s & 15.0s & 15.0s & 15.0s & 15.0s & 15.0s & 15.0s\\ 
			\toprule
			SQL001 & $\times$ & $\times$ & $\times$ & $\times$ & $\times$ & $\times$ & $\times$ & $\times$\\ 
			SQL002 & 5.1h & 23.0m & 3.0m & 28.0m & 25.0s & 1.0m & 1.6h & 4.0h\\ 
			SQL003 & $\times$ & $\times$ & 10.9h & $\times$ & $\times$ & $\times$ & $\times$ & 9.0h\\ 
			SQL004 & $\times$ & $\times$ & $\times$ & $\times$ & $\times$ & $\times$ & $\times$ & $\times$\\ 
			SQL005 & $\times$ & $\times$ & $\times$ & $\times$ & $\times$ & $\times$ & $\times$ & 18.2h\\ 
			SQL006 & $\times$ & $\times$ & 1.0h & 2.2h & $\times$ & $\times$ & 4.0h & $\times$\\ 
			SQL007 & 51.0m & 31.0m & 3.0m & 1.0h & 20.0s & 60.0s & 59.1m & 1.0h\\ 
			SQL008 & $\times$ & $\times$ & $\times$ & $\times$ & $\times$ & $\times$ & $\times$ & $\times$\\ 
			SQL009 & $\times$ & 8.7h & 37.0m & 18.0h & 2.2h & 38.0m & 10.2h & 13.0m\\ 
			SQL010 & 12.6h & 1.0h & 12.0m & 1.0h & 25.0s & 40.0m & 2.0h & 3.1h\\ 
			SQL011 & $\times$ & $\times$ & 12.8h & $\times$ & $\times$ & $\times$ & 13.5h & 20.0h\\ 
			SQL012 & $\times$ & $\times$ & $\times$ & $\times$ & $\times$ & $\times$ & $\times$ & 3.2h\\ 
			SQL013 & $\times$ & $\times$ & 18.0h & $\times$ & 3.0h & $\times$ & 20.1h & $\times$\\ 
			SQL014 & $\times$ & $\times$ & 40.0m & 20.9h & 1.0h & $\times$ & 2.3h & 30.0m\\ 
			SQL015 & 5.0h & 23.0m & 3.0m & 1.1h & 25.0s & 1.0m & 58.0m & 1.0h\\ 
			SQL016 & 6.5h & 1.0h & 5.0m & 57.0m & 25.0s & 3.0m & 31.0m & 20.0s\\ 
			\bottomrule
		\end{tabular}
\end{table*}

\end{document}